%% file: arxiv2.tex
\documentclass[12pt,a4paper]{article}
\usepackage{tikz}
\usepackage{pgf}

\usepackage{amsmath}
\usepackage{booktabs}
\usepackage{cite}
\usepackage[colorlinks=true,linkcolor=black, citecolor=black,urlcolor=black]{hyperref}

\usepackage{multirow,graphics}
\usepackage{enumerate}
\usepackage{amstext}
\usepackage{amssymb}
\usepackage{amsmath}

\usepackage{url}
\usepackage{graphicx}
\usepackage{subcaption}
\usepackage{color}

\usepackage[nodisplayskipstretch]{setspace}
\usepackage{tcolorbox}
\usepackage{float}

\usepackage{wasysym}
\usepackage{ytableau}
\usepackage{textcomp}

\usetikzlibrary{decorations.pathmorphing}
\usetikzlibrary{decorations.markings}
\usetikzlibrary{intersections}
\usetikzlibrary{calc}
\usetikzlibrary{positioning}
\usetikzlibrary{3d}
\usetikzlibrary{shapes}
\usetikzlibrary{angles}
\usetikzlibrary{fit}
\usetikzlibrary{backgrounds}

\tikzset{
photon/.style={decorate, decoration={snake}},
particle/.style={postaction={decorate},
    decoration={markings,mark=at position .5 with {\arrow{>}}}},
antiparticle/.style={postaction={decorate},
    decoration={markings,mark=at position .5 with {\arrow{<}}}},
gluon/.style={decorate, decoration={coil,amplitude=2pt, segment length=4pt},color=purple},
wilson/.style={color=blue, thick},
scalarZ/.style={postaction={decorate},decoration={markings, mark=at position .5 with{\arrow[scale=1]{stealth}}}},
scalarX/.style={postaction={decorate}, dashed, dash pattern = on 4pt off 2pt, dash phase = 2pt, decoration={markings, mark=at position .53 with{\arrow[scale=1]{stealth}}}},
scalarZw/.style={postaction={decorate},decoration={markings, mark=at position .75 with{\arrow[scale=1]{stealth}}}},
scalarXw/.style={postaction={decorate}, dashed, dash pattern = on 4pt off 2pt, dash phase = 2pt, decoration={markings, mark=at position .60 with{\arrow[scale=1]{stealth}}}},
frozen/.style={inner sep=0.7mm, rectangle,draw},
frozenblue/.style={rectangle, draw, fill=blue!20, inner sep=0.7mm},
norm/.style={->, draw, shorten <=2pt, shorten >=2pt},
diag/.style={->, draw, shorten <=5pt, shorten >=3pt},
every node/.style={inner sep=0.5mm},
webarrow/.style={postaction={decorate},,decoration={markings, mark=at position .5 with{\arrow[scale=1]{stealth}}}}
}

\makeatletter
\tikzoption{canvas is plane}[]{\@setOxy#1}
\def\@setOxy O(#1,#2,#3)x(#4,#5,#6)y(#7,#8,#9)%
  {\def\tikz@plane@origin{\pgfpointxyz{#1}{#2}{#3}}%
   \def\tikz@plane@x{\pgfpointxyz{#4}{#5}{#6}}%
   \def\tikz@plane@y{\pgfpointxyz{#7}{#8}{#9}}%
   \tikz@canvas@is@plane
  }
\makeatother

\newcommand{\Gr}[1]{$\operatorname{Gr}(#1)$}
\newcommand{\Dr}[1]{$\operatorname{Dr}(#1)$}
\newcommand{\pGr}[1]{$\operatorname{Gr}^+(#1)$}

\newcommand{\TGr}[1]{$\operatorname{Tr}(#1)$}
\newcommand{\TpGr}[1]{$\operatorname{Tr}^+(#1)$}

\newcommand\be{\begin{equation}}
\newcommand\ee{\end{equation}}

\newcommand{\cA}{\mathcal{A}}

\normalsize
\makeatletter
\divide\@tempdima\baselineskip
\@tempcnta=\@tempdima
\skip\@mpfootins = \skip\footins
\renewcommand{\@dotsep}{10000}
\makeatother

\begin{document}
\numberwithin{equation}{section}
\begin{center}
\phantom{vv}

\vspace{3cm}
\bigskip

{\Large \bf Tropical Grassmannians, cluster algebras and scattering amplitudes}

\bigskip
{\mbox {\bf James Drummond, Jack Foster,  \"Omer G\"urdo\u gan, Chrysostomos Kalousios}}%
\footnote{ {\sffamily \{\tt j.a.foster, j.m.drummond, o.c.gurdogan, c.kalousios\}@soton.ac.uk }}
\bigskip

{\em School of Physics \& Astronomy, University of Southampton,\\
  Highfield, Southampton, SO17 1BJ, United Kingdom.}

\vspace{3cm} \bigskip \vspace{30pt} {\bf Abstract}
\end{center}

\noindent We provide a cluster-algebraic approach to the computation
of the recently introduced generalised biadjoint scalar amplitudes related to Grassmannians ${\rm Gr}(k,n)$. 
A finite cluster algebra provides a natural triangulation for the tropical Grassmannian
whose volume computes the scattering amplitudes. Using this method
one can construct the entire colour-ordered amplitude via mutations starting from a
single term.

\newpage
\phantom{vv}
\vspace{1cm}
\hrule
\tableofcontents

\bigskip
\medskip

\hrule
\newpage
\section{Introduction}

Recently a very interesting connection between scattering amplitudes
and tropical geometry has been uncovered
\cite{Cachazo:2019ngv,Cachazo:2019apa}. The connection outlined so far
is for tree-level biadjoint $\phi^3$ amplitudes, which can be related
to the series of tropical Grassmannians ${\rm Gr}(2,n)$ and a
generalisation to higher Grassmannians ${\rm Gr}(k,n)$. Such
amplitudes also have a formulation in terms of a set of scattering
equations which generalise the usual scattering equations of
\cite{Cachazo:2013gna,Cachazo:2013hca,Cachazo:2013iea}

Tropical Grassmannians are defined as a space of solutions to a set of
tropical hypersurface conditions which derive from the defining
Pl\"ucker relations of the Grassmannian.  An important ingredient in
the relation to the generalised biadjoint scattering amplitudes is the
notion of positivity which singles out a particular region in the
tropical Grassmannian. We describe here the tropical formulation of
the Grassmannian spaces and how to select the positive region. We will
see that this coincides with the criteria recently used in
\cite{Cachazo:2019ngv,Cachazo:2019apa} to determine the generalised
$\phi^3$ amplitudes for \Gr{3,6} and \Gr{3,7}.

We will also develop the link further and describe a relation of the
positive tropical Grassmannians to certain cluster algebras, as
developed by Fomin and Zelevinsky \cite{1054.17024,1021.16017}. These
same cluster algebras have also arisen in the study of the
singularities of \emph{loop} amplitudes in planar $\mathcal{N}=4$
super Yang-Mills theory \cite{Golden:2013xva}. The cluster algebra
picture provides extremely efficient calculational tools for
determining the relevant positive solutions to the tropical
hypersurface conditions, allowing for spaces of even quite large
dimension to be simply constructed.

Once the positive region is obtained, the generalised biadjoint
$\phi^3$ amplitudes can be constructed as its volume in a direct
generalisation of the picture described in
\cite{Arkani-Hamed:2017mur}. Such a volume can be obtained additively
via a triangulation of the region into simplexes. One such
triangulation is provided by the (dual of the) associated cluster
polytope. For the ${\rm Gr}(2,n)$ cases these polytopes are the
$A_{n-3}$ associahedra. In the \Gr{3,6} case this corresponds to the
$D_4$ polytope while in the \Gr{3,7} case it is the $E_6$ polytope
familiar from heptagon amplitude in planar $\mathcal{N}=4$ super
Yang-Mills theory. For the ${\rm Gr}(3,8)$ case we can obtain a
triangulation from the $E_8$ cluster polytope. This triangulation has
the feature that it makes use of eight spurious vertices generated by
the cluster algebra but not strictly needed to compute the volume. The
above cases exhaust the list of finite Grassmannian cluster algebras.

A feature of the polytopes arising as positive tropical Grassmannians
is that in general their facets are not all simplexes. This means that
there is a redundancy in parametrising their volumes since they may be
triangulated (or cut into simplexes) in multiple ways, each yielding a
seemingly different but actually equivalent way of obtaining the
volume. In physical language this means there are multiple ways of
writing the amplitude which are in fact equivalent due to non-trivial
identities between different contributions.

The non-simplicial nature of certain facets may also have a bearing on
the analytic structure of loop amplitudes in planar $\mathcal{N}=4$
theory. In the case of \Gr{3,7} it would be relevant for the heptagon amplitudes studied in \cite{Drummond:2014ffa,Drummond:2017ssj} where it should be related to the recently discovered property of cluster adjacency
\cite{Drummond:2017ssj,Drummond:2018dfd} which forbids certain consecutive pairs of
branch cuts in loop amplitudes and is related to the Steinmann
relations. The non-simplicial facets can be thought of as a
combination of simplexes, which corresponds in the cluster polytope to
shrinking edges so that many clusters combine together.

We describe how the cases of \Gr{3,6} and \Gr{3,7}
fit into the above picture and we extend it to the case of \Gr{3,8}
whose positive tropical version corresponds to the $E_8$ cluster
algebra.
Since all these cluster algebras are finite, the triangulation
procedure works in exactly the same way for all of them. Nevertheless
the correspondence of between the cluster algebra and the fan for each
case contains intricacies of different nature with valuable lessons
and we elaborate on these in sections dedicated to different
Grassmannians.

Before this we review the interpretation of the biadjoint $\phi$ amplitude as the volume of the dual to a kinematic realisation of the associahedron. We then illustrate all the main principles of the tropical
Grassmannian, its positive part and the connection to cluster algebras
in the example of \Gr{2,5}.

\section{Amplitudes from volumes of dual associahedra}

In \cite{Arkani-Hamed:2017mur} a connection between biadjoint scalar amplitudes and volumes was made. The main idea is to introduce a kinematic realisation of the associahedron. This is done as follows. Given an ordered set of light-like momenta $p_1,\ldots,p_n$ satisfying momentum conservation one introduces dual coordinates,
\be
x_{i+1}-x_i = p_i\,,
\ee
with all indices treated modulo $n$.
The $\frac{1}{2}n(n-3)$ square distances $(x_i-x_j)^2= X_{ij}$ can be related to Mandelstam invariants via
\be
X_{ij} = s_{i,i+1,\ldots,j-1} = (p_i +p_{i+1} + \ldots p_{j-1})^2\,.
\ee
Note that the momenta being null implies $X_{i,i+1}=0$.
The two-particle Mandelstam invariants $s_{ij} = (p_i+p_j)^2$ can be related to the dual variables via
\be
s_{ij} = X_{i,j+1} + X_{i+1,j} - X_{ij} - X_{i+1,j+1}\,.
\ee

To define the \emph{kinematic associahedron} we take all $X_{ij}$ positive and choose $(n-3)$ coordinates, e.g. the $X_{1,i}$ for $i=3,\ldots,(n-1)$. The remaining $\frac{1}{2}(n-2)(n-3)$ independent variables need to be constrained in order to obtain a space of dimension $(n-3)$. To do this we impose $\frac{1}{2}(n-2)(n-3)$ conditions which we take to be of the form 
\be
\label{kinematic_conds}
s_{ij} = - c_{ij}\,, \qquad 2\leq i<j \leq n\,, \quad i\leq j-2\,,
\ee
for \emph{positive} constants $c_{ij}$. The coordinates $X_{1,i}$ are then constrained to run only over a certain region: the kinematic associahedron.

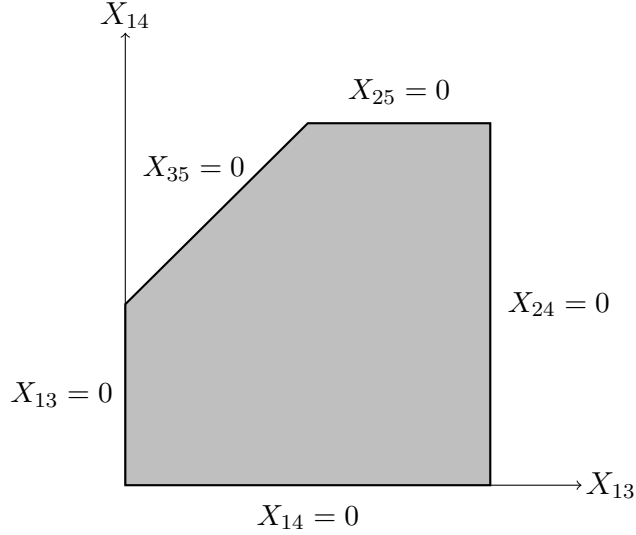
\begin{figure}
\centering
\begin{tikzpicture}[scale=1.2]
\draw[->] (0,0) -- (0,5) node[above]{$X_{14}$};
\draw[->] (0,0) -- (5,0) node[right]{$X_{13}$};
\draw[thick,black,fill=lightgray] (0,0) -- (4,0) -- (4,4) -- (2,4) -- (0,2) -- cycle;
\node[] at (2,-0.35) {\small $X_{14}=0$};
\node[] at (-0.7,1) {\small $X_{13}=0$};
\node[] at (0.75,3.5) {\small $X_{35}=0$};
\node[] at (3,4.35) {\small $X_{25}=0$};
\node[] at (4.75,2) {\small $X_{24}=0$};
\end{tikzpicture}
\caption{The shaded area is the kinematic associahedron for $n=5$.}
\label{KA25}
\end{figure}

For the $n=5$ example the conditions (\ref{kinematic_conds}) become
\begin{align}
X_{35} &= c_{35} + X_{13} - X_{14}\,, \notag \\
X_{25} &= c_{25} + c_{35} - X_{14}\, \notag \\
X_{24} &= c_{24} + c_{25} - X_{13}\,. 
\end{align}
The coordinates $(X_{13},X_{14})$ then run over a region with the shape of a pentagon as shown in Fig. \ref{KA25}.

To obtain the dual of the kinematic associahedron it is helpful to embed it into projective space $\mathbb{P}^{\,n-3}$. We introduce the auxiliary point $Y=(1,X_{13},X_{14},\ldots,X_{1,n-1})$.
The boundary conditions $X_{ij}=0$ of the kinematic associahedron become $Y \cdot W_{ij} = 0$ with $W_{ij}$ given by projective dual vectors determined by the conditions (\ref{kinematic_conds}).

In the case $n=5$ we have $Y=(1,X_{13},X_{14})$ and 
\begin{align}
W_{13} &= (0,1,0) \,, \notag \\
W_{14} &= (0,0,1) \,, \notag \\
W_{24} &= (c_{24} + c_{25},-1,0)\,, \notag \\
W_{25} &= (c_{25} + c_{35},0,-1)\,, \notag \\
W_{35} &= (c_{35}, 1 , -1)\,.
\label{G25dualvectors}
\end{align}
These dual vectors define the dual to the ${\rm Gr}(2,5)$ kinematic associahedron. Its volume may be computed by first triangulating it, e.g. by picking the reference point $W_*=(1,0,0)$ and adding the volume of the five triangles formed by $W_*$ and two adjacent dual vectors according to
\be
{\rm Vol}(W_1,W_2,W_3) = \frac{\langle W_1 W_2 W_3\rangle}{(Y\cdot W_1) (Y \cdot W_2)(Y \cdot W_3)}\,.
\ee

In this way we obtain the sum of five terms,
\begin{align}
{\rm Vol}(\mathcal{A}^*) 
&= \frac{1}{X_{13}X_{14}} + \frac{1}{X_{14}X_{24}} + \frac{1}{X_{24}X_{25}} + \frac{1}{X_{25}X_{35}} + \frac{1}{X_{35}X_{13}}\,, \notag \\
&= \frac{1}{s_{12} s_{45}} + \frac{1}{s_{45} s_{23}} + \frac{1}{s_{23} s_{15}} + \frac{1}{s_{15} s_{34}} + \frac{1}{s_{34} s_{12}}
\end{align}
and we recognise the obtained representation as the Feynman diagram expansion for the canonically ordered biadjoint $\phi^3$ amplitude.

\section{Tropical Grassmannians and amplitudes} 
\label{sect-tropical}

The Grassmannian \Gr{k,n} is the space of $k$-planes in $n$
dimensions. The Grassmannian can therefore be parametrised
by a $k \times n$ complex matrix with the $k$ rows specifying a $k$
plane. Since the plane is invariant under the action of $GL(k)$
transformations one must mod out by the action of $GL(k)$, leaving a
space of dimension $k(n-k)$.

The Grassmannian may also be specified in terms of the minors of the
matrix. The $(k \times k)$ minors $p_{i_1,\ldots,i_k}$ (Pl\"ucker
coordinates) of any matrix obey homogeneous quadratic relations
(Pl\"ucker relations) obtained by antisymmetrising $(k+1)$ indices,
\be p_{i_1,\ldots,i_r,[i_{r+1},\ldots
  i_k}p_{j_1,\ldots,j_{r+1}],j_{r+2},\ldots,j_k} = 0\,.  \ee In the
\Gr{2,n} case the Pl\"ucker relations are given by the familiar
$\binom{n}{4}$ three-term equations
\begin{equation}\label{plucker2n}
p_{ij}p_{kl}-p_{ik}p_{jl}+p_{il}p_{jk} = 0\,, \qquad 1 \leq i < j < k < l  \leq n\,.
\end{equation}

The Pl\"ucker relations define a subspace in the Pl\"ucker space
parametrised by the $\binom{n}{k}$ Pl\"ucker coordinates
$p_{i_1,\ldots,i_k}$. Algebraically this space may be thought of as
the ideal generated by the quadratic Pl\"ucker relations inside the
ring of polynomials in the Pl\"ucker coordinates. After quotienting by
a global rescaling of all Pl\"ucker coordinates the subspace
satisfying the Pl\"ucker relations can be identified with the
Grassmannian \Gr{k,n} of dimension $k(n-k)$.

The original Pl\"ucker relations are actually homogeneous in $n$
independent rescalings
$p_{i_1,\ldots,i_k} \rightarrow t_{i_1} \ldots t_{i_k}
p_{i_1,\ldots,i_k}$ with $t_i \in \mathbb{C}^*$. If we quotient by all
of these scalings instead of just the overall scaling we obtain a
smaller space, \be {\rm Conf}_n(\mathbb{P}^{k-1}) = {\rm
  Gr}(k,n)/(\mathbb{C}^*)^{n-1}\,, \ee which has dimension
$m=(k-1)(n-k-1)$ and corresponds to taking the columns of our original
$(k \times n)$ to be elements of $\mathbb{P}^{k-1}$ instead of
$\mathbb{C}^k$.

There exists a \emph{tropical} version of the above construction. In
tropical geometry one takes the generating relations of the ideal and
replaces multiplication with addition and addition with minimum. For
example the generating quadratic polynomials of the \Gr{2,n} Pl\"ucker
relations (\ref{plucker2n}) become the tropical polynomials
\begin{equation}
\label{troppoly}
{\rm min}(w_{ij}+w_{kl}, w_{ik}+w_{jl} , w_{il}+w_{jk})\,,
\end{equation}
which are piecewise linear maps on the space of $\binom{n}{2}$
variables $w_{ij}\in \mathbb{R} $. 

Piecewise linear maps have special surfaces between one
region of linearity and another. Such surfaces are called tropical
hypersurfaces and are attained when at least two of the terms of the
tropical polynomial simultaneously attain the minimum. In other words
the tropical polynomial (\ref{troppoly}) defines the following
tropical hypersurfaces,
\begin{equation}\label{tropical_conditions}
\begin{aligned}
w_{ij}+w_{kl}&=w_{ik}+w_{jl} \leq w_{il}+w_{jk} \\
\text{or}\quad w_{ij}+w_{kl} &= w_{il}+w_{jk} \leq w_{ik}+w_{jl} \\
\text{or}\quad w_{ik}+w_{jl} &= w_{il}+w_{jk} \leq w_{ij}+w_{kl}\,.
\end{aligned}
\end{equation}

When we have many polynomial relations we must simultaneously satisfy
the conditions arising from each polynomial relation. In the case of
\Gr{2,n} we must simultaneously satisfy the hypersurface relations
coming from every Pl\"ucker relation, i.e. for every choice of
$\{i,j,k,l\}$ in (\ref{plucker2n}).

Note that for any solution $\{w_{ij}\}$, any global \emph{positive}
rescaling of the $w_{ij}$ will also obey the conditions. Solutions
therefore form rays emanating from the origin and can be represented
by an $\binom{n}{2}$-component vector, or more generally for \Gr{k,n}
an $\binom{n}{k}$-component vector.  Note also that if $\{w_{ij} \}$
are solutions of the above conditions then so are
$\{w_{ij} + a_i + a_j\}$ for any set of $n$ constants
$a_i \in \mathbb{R}$. Such a shift symmetry is referred to as
\emph{lineality}. In the context of generalised biadjoint scattering amplitudes it corresponds to momentum conservation.

Quotienting the space of solutions of the tropical hypersurface conditions (\ref{tropical_conditions}) by a single global shift with $a_i=a$ corresponds to the tropical version of the Grassmannian. Quotienting by all shifts corresponds to the tropical version of the space ${\rm Conf}_n(\mathbb{P}^{k-1})$. Here we are interested in the latter case where we quotient by all shifts. Despite this we will refer to the space obtained simply as the tropical Grassmannian and we use the notation \TGr{k,n} to denote it.

The sign of the individual terms of the Pl\"ucker relations
\eqref{plucker2n} is lost through tropicalisation. We
can recover the information by identifying \emph{positive} hypersurfaces as those whose
defining terms in \eqref{plucker2n} have \emph{opposite} signs \cite{2015arXiv151102699B}. This prescription defines the \emph{positive tropical Grassmannian}.
The positive part of \TGr{2,n} (denoted \TpGr{2,n}) is closely related to the dual of the kinematic associahedron that we described above and hence can be identified with the canonically ordered amplitude of the bi-adjoint $\phi^3$ theory.  This fact is at the heart of the recent generalisation of the biadjoint amplitudes to general ${\rm Tr}(k,n)$ \cite{Cachazo:2019ngv}. In Sect. \ref{webs} we give a more detailed introduction to the positive tropical Grassmannian following \cite{2003math.....12297S}.

Such generalised biadjoint amplitudes can also be related to a generalisation of
the scattering equations
\cite{Cachazo:2013gna,Cachazo:2013hca,Cachazo:2019ngv} to
$\mathbb{CP}^{k-1}$ and through them to amplitudes of a generalised scalar
bi-adjoint theory \cite{Cachazo:2013iea}.  Focusing for simplicity to
$k=3$, we consider homogenous coordinates of $n$ particles on
$\mathbb{CP}^{2}$ and form the $3\times n$ matrix
\begin{equation}
m=
\left(
\begin{array}{cccc}
 1 & 1 & \cdots & 1 \\
 x_1 & x_2 & \cdots & x_n \\
 y_1 & y_2 & \cdots & y_n \\
\end{array}
\right).
\end{equation}
We then define the potential function 
\begin{equation}
S_3 = \sum_{1\leq i < j < k \leq n} s_{ijk} \log [ ijk ], 
\end{equation}
where $[ ijk ]$ represent minors of $m$ and $s_{ijk}$ are generalized Mandelstam variables that satisfy $\sum_{j\neq k} s_{ijk} =0,~ \forall i$.  We can now write down the amplitude of a generalised scalar theory as
\begin{equation}\label{A_n}
A^{(3)}_n(\alpha | \beta) =  \frac{1}{\text{vol} (\text{SL}(3,\mathbb{C}))} 
\int \prod_i \mathrm{d} x_i \mathrm{d} y_i \delta(S_{3,x_i}) \delta (S_{3,y_i})
\text{PT}(\alpha) \text{PT}(\beta),
\end{equation}
where $S_3,i$ denotes derivative with respect to $i$ and the
generalized Parke-Taylor factors involve two orderings $\alpha$ and
$\beta$ and are given by
\begin{equation}
\text{PT} (\mathbb{I}) = \frac{1}{[123][234] \cdots [n12]}.
\end{equation}
The positive region of the tropical computation should then equal
\eqref{A_n} for the canonical ordering $\alpha=\beta=\mathbb{I}$.

Let us consider explicit examples of the tropical Grassmannian \cite{SpeyerSturmfels}. The simplest case is \Gr{2,4}, defined by the single Pl\"ucker relation, 
\be
\label{Gr24Plucker}
p_{12}p_{34}-p_{13}p_{24}+p_{14}p_{23} = 0\,.  \ee In this case the
tropical hypersurface conditions have three solutions (modulo
lineality), given by the three possibilities in
(\ref{tropical_conditions}) with $\{i,j,k,l\} = \{1,2,3,4\}$. They are
represented by the following six component vectors corresponding to
the canonical ordering of the
$\{w_{12},w_{13},w_{14},w_{23},w_{24},w_{34}\}$,
\begin{align}
e_{12} &= (1,0,0,0,0,0)\,, \notag \\
e_{13} &= (0,1,0,0,0,0)\,, \notag \\
e_{14} &= (0,0,1,0,0,0)\, .
\end{align}
Of these only the first and third are positive. Note that one may not
generally add solutions to obtain other solutions, the above vectors
represent three distinct solutions. Note also that because of the
shift symmetry $w_{ij} \mapsto w_{ij} + a_i + a_j$ the following
vectors
\begin{align}
e_{34} &= (0,0,0,0,0,1)\,, \notag \\
e_{24} &= (0,0,0,0,1,0)\,, \notag \\
e_{23} &= (0,0,0,1,0,0)\,
\end{align}
are equivalent to the original three. This shift symmetry has the
interpretation of momentum conservation once the solution vectors
$e_{ij}$ are contracted with a canonically ordered vector of Mandelstam invariants $y=(s_{12},\ldots,s_{34})$ entering the massless biadjoint scattering amplitudes.

Let us now describe the \Gr{2,5} case.  In this case we have
ten Pl\"ucker coordinates $p_{ij}$ and the Pl\"ucker relations are
given by (\ref{Gr24Plucker}) and four more relations obtained from
cyclic rotation of the labels. These relations give rise to the
tropical hypersurface conditions (\ref{tropical_conditions}) for
$\{i,j,k,l\}$ given by $\{1,2,3,4\}$, $\{1,2,3,5\}$, $\{1,2,4,5\}$,
$\{1,3,4,5\}$ and $\{2,3,4,5\}$. Each of these five cases must be
simultaneously satisfied.

We arrange the coordinates in the standard, lexicographical order,
\be
\{w_{12},w_{13},w_{14},w_{15},w_{23},w_{24},w_{25},w_{34},w_{35},w_{45}\}
\ee
and define ray vectors as
\begin{align}
e_{12} &= (1,0,0,0,0,0,0,0,0,0)\,,\notag \\
e_{13} &= (0,1,0,0,0,0,0,0,0,0)\,, \notag \\
&\,\,\, \vdots\notag\\
e_{45} &= (0,0,0,0,0,0,0,0,0,1)
\label{G25rays}
\end{align}
and so on. The vectors $e_{ij}$ so defined are simultaneously solutions to all five of the tropical hypersurface conditions.

In this case we can also combine certain solutions. For example we find that any positive linear combination $a e_{12} + b e_{34}$ with $a,b>0$ is also a solution. However no positive linear combination $a e_{12} + b e_{13}$ is a solution. We thus obtain a notion of \emph{connectivity} of solutions: two solutions are connected if any positive linear combination of them is a solution. We say that there is an \emph{edge} between such solutions. In the case of ${\rm Gr}(2,5)$ we can never combine three or more solutions to obtain another solution. In higher dimensional examples one can obtain triangles of solutions and higher dimensional faces. 

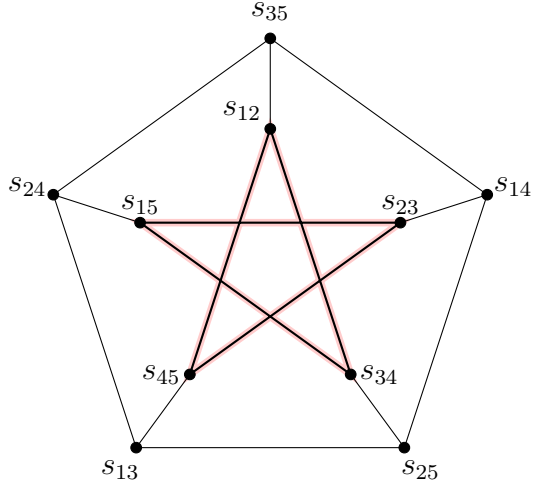
\begin{figure}
\begin{center}
  \begin{tikzpicture}[label distance = 0.18]
    \pgfmathsetmacro{\innerrad}{1.8};
    
    \coordinate (i1) at (90:\innerrad);
    \coordinate (i2) at (234:\innerrad);
    \coordinate (i3) at (18:\innerrad);
    \coordinate (i4) at (162:\innerrad);
    \coordinate (i5) at (-54:\innerrad);

    \coordinate (o1) at (90:3);
    \coordinate (o2) at (234:3);
    \coordinate (o3) at (18:3);
    \coordinate (o4) at (162:3);
    \coordinate (o5) at (-54:3);

    \node[label=above left:$s_{12}$] at (i1) {};
    \node[label=left:$s_{45}$] at (i2) {};
    \node[label=above:$s_{23}$] at (i3) {};
    \node[label=above:$s_{15}$] at (i4) {};
    \node[label=right:$s_{34}$] at (i5) {};

    \node[] at ($(o1)+0.12*(o1)$) {$s_{35}$};
    \node[] at ($(o2)+0.12*(o2)$) {$s_{13}$};
    \node[] at ($(o3)+0.12*(o3)$) {$s_{14}$};
    \node[] at ($(o4)+0.12*(o4)$) {$s_{24}$};
    \node[] at ($(o5)+0.12*(o5)$) {$s_{25}$};

    \draw[line width=1mm,opacity=0.2,red] (i1) -- (i2) -- (i3) -- (i4) -- (i5) -- cycle;
    \draw[thick,black] (i1) -- (i2) -- (i3) -- (i4) -- (i5) -- cycle;
    \draw (o1) -- (o3) -- (o5) -- (o2) -- (o4) -- cycle;
    \foreach \i in {1,...,5}{
      \draw (i\i) -- (o\i);
    }
    \foreach \i in {1,...,5}{
      \draw[fill=black] (i\i) circle (0.07);
      \draw[fill=black] (o\i) circle (0.07);
    }

  \end{tikzpicture}
\caption{The 10 vertices and 15 edges of the full ${\rm Tr}(2,5)$ space.  The highlighted star is the positive region.  It corresponds to the canonical order amplitude of the scalar bi-adjoint $\phi^3$ theory. }
\label{Petersen_graph}
\end{center}
\end{figure}

Performing permutations on the indices leads us to find 15 edges
between the 10 vertices given by the $e_{ij}$. The full set of
solutions corresponding to the tropical Grassmannian \TGr{2,5} can be
depicted by the Petersen graph shown in Fig. \ref{Petersen_graph}.

The positive part ${\rm Tr}^+(2,5)$ is identified with those solutions where only the first and third possibilities in (\ref{tropical_conditions}) are allowed in each of the five cases. This picks out the solutions $\{e_{12},e_{23},e_{34},e_{45},e_{15}\}$. The positive part is then given by the positive rays and the edges between them (any positive linear combination of connected positive solutions is a positive solution). The positive part is highlighted in Fig. \ref{Petersen_graph}.

\section{The positive tropical Grassmannian from webs}
\label{webs}

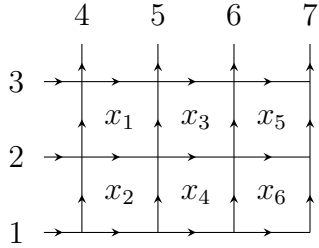
\begin{figure}
\begin{center}
  \begin{tikzpicture}[node distance=2cm]
    \foreach \i in {1,...,3}{
      \draw[webarrow] (0.5,{\i-1}) node[left=2mm]{\i}--(1,{\i-1});
    }
    
    \foreach \i [evaluate=\i as \lab using {int(\i+3)}] in {1,...,4}{
      \draw[webarrow] (\i,2) --(\i,2.5) node[above=2mm]{{\lab}};
    }
    \foreach \i in {1,...,4}{
      \foreach \j in {0,...,1}{
        \draw[webarrow] (\i,\j) --(\i,{\j+1});
        }
    }
    \foreach \i in {1,...,3}{
      \foreach \j in {0,...,2}{
        \draw[webarrow] ({\i},\j) --({\i+1},\j);
        }
    }

    \node at (1.5,1.5) {$x_1$};
    \node at (2.5,1.5) {$x_3$};
    \node at (3.5,1.5) {$x_5$};
    \node at (1.5,0.5) {$x_2$};
    \node at (2.5,0.5) {$x_4$}; 
    \node at (3.5,0.5) {$x_6$};
  \end{tikzpicture}
  \caption{Example web diagram for ${\rm Gr}(3,7)$. }
   \label{webex}
  \end{center}
\end{figure}

In \cite{2003math.....12297S} an alternative way of describing just
the positive part \TpGr{k,n} was given. In this approach one
introduces a grid called a \emph{web diagram} with labels
$\{1,\ldots,k\}$ on the horizontal edge and labels
$\{(k+1),\ldots,n\}$ on the vertical edge. The squares of the grid are
populated with variables $x_i$. In Fig. \ref{webex} we illustrate the
general procedure in the case of \TGr{3,7}.

A Pl\"ucker coordinate is indexed by a set $K$ of $k$ distinct labels chosen from $\{1,\ldots,n\}$. We denote the set $\{1,\ldots,k\}$ by $[k]$. We may then associate a Pl\"ucker coordinate $p_{K}$ to a set of paths on the web diagram as follows. Consider sets $S$ of non-intersecting paths consistent with the arrows which go from $[k]\setminus([k]\cap K)$ to $K\setminus ([k]\cap K)$. We denote the set of all such sets as ${\rm Path}(K)$. For each path in a given set $S$ we record the product of the variables in the squares above the path (if there are no squares above the path we record the value 1). For a set $S$ of paths we take the product over all paths in the set which we denote by ${\rm Prod}_S(x)$ (if the set is empty we record the value 1). Finally we sum over all possible choices of sets $S$ of such non-intersecting paths, i.e. we sum over $S\in {\rm Path}(K)$,
\be
\label{webPlucker}
p_K = \sum_{S \in {\rm Path}(K)} {\rm Prod}_S(x)\,.
\ee 

\begin{figure}
\begin{center}
\begin{subfigure}{0.3\textwidth}
  \begin{tikzpicture}[node distance=2cm]
    \draw[gray] (0.5,2) node[left]{3}--(4,2);
    \draw[gray] (0.5,1) node[left]{2}--(4,1);
    \draw[gray] (0.5,0) node[left]{1}--(4,0);

    \draw[gray] (1,0) --(1,2.5) node[above]{4};
    \draw[gray] (2,0)--(2,2.5) node[above]{5};
    \draw[gray] (3,0)--(3,2.5) node[above]{6};
    \draw[gray] (4,0)--(4,2.5) node[above]{7};

    \node at (1.5,1.5) {$x_1$};
    \node at (2.5,1.5) {$x_3$};
    \node at (3.5,1.5) {$x_5$};
    \node at (1.5,0.5) {$x_2$};
    \node at (2.5,0.5) {$x_4$}; 
    \node at (3.5,0.5) {$x_6$};
    
    \draw[webarrow,thick] (0.5,1) -- (1,1);
    \draw[webarrow,thick] (1,1) -- (1,2);
    \draw[webarrow,thick] (1,2) -- (3,2);
    \draw[webarrow,thick] (3,2)  -- (3,2.5);
    \draw[webarrow,thick] (0.5,0) -- (2,0);
    \draw[webarrow,thick] (2,0) -- (2,1);
    \draw[webarrow,thick] (2,1) -- (4,1);
    \draw[webarrow,thick] (4,1)-- (4,2.5);
  \end{tikzpicture}
\end{subfigure}  
\begin{subfigure}{0.3\textwidth}
 \begin{tikzpicture}[node distance=2cm]
    \draw[gray] (0.5,2) node[left]{3}--(4,2);
    \draw[gray] (0.5,1) node[left]{2}--(4,1);
    \draw[gray] (0.5,0) node[left]{1}--(4,0);

    \draw[gray] (1,0) --(1,2.5) node[above]{4};
    \draw[gray] (2,0)--(2,2.5) node[above]{5};
    \draw[gray] (3,0)--(3,2.5) node[above]{6};
    \draw[gray] (4,0)--(4,2.5) node[above]{7};

    \node at (1.5,1.5) {$x_1$};
    \node at (2.5,1.5) {$x_3$};
    \node at (3.5,1.5) {$x_5$};
    \node at (1.5,0.5) {$x_2$};
    \node at (2.5,0.5) {$x_4$}; 
    \node at (3.5,0.5) {$x_6$};
    
    \draw[webarrow,thick] (0.5,1) -- (1,1);
    \draw[webarrow,thick] (1,1) -- (1,2);
    \draw[webarrow,thick] (1,2)-- (3,2);
    \draw[webarrow,thick] (3,2)-- (3,2.5);
    \draw[webarrow,thick] (0.5,0) -- (3,0);
    \draw[webarrow,thick] (3,0) -- (3,1);
    \draw[webarrow,thick] (3,1) -- (4,1);
    \draw[webarrow,thick] (4,1) -- (4,2.5);
  \end{tikzpicture}  
  \end{subfigure}  
  \begin{subfigure}{0.3\textwidth}
 \begin{tikzpicture}[node distance=2cm]
    \draw[gray] (0.5,2) node[left]{3}--(4,2);
    \draw[gray] (0.5,1) node[left]{2}--(4,1);
    \draw[gray] (0.5,0) node[left]{1}--(4,0);

    \draw[gray] (1,0) --(1,2.5) node[above]{4};
    \draw[gray] (2,0)--(2,2.5) node[above]{5};
    \draw[gray] (3,0)--(3,2.5) node[above]{6};
    \draw[gray] (4,0)--(4,2.5) node[above]{7};

    \node at (1.5,1.5) {$x_1$};
    \node at (2.5,1.5) {$x_3$};
    \node at (3.5,1.5) {$x_5$};
    \node at (1.5,0.5) {$x_2$};
    \node at (2.5,0.5) {$x_4$}; 
    \node at (3.5,0.5) {$x_6$};
    
    \draw[webarrow,thick] (0.5,1) -- (1,1);
    \draw[webarrow,thick] (1,1) -- (1,2);
    \draw[webarrow,thick] (1,2) -- (3,2);
    \draw[webarrow,thick] (3,2) -- (3,2.5);
    \draw[webarrow,thick] (0.5,0) -- (4,0);
    \draw[webarrow,thick] (4,0) -- (4,2.5);
  \end{tikzpicture}  
 \end{subfigure}  
 \\
   \begin{subfigure}{0.3\textwidth}
 \begin{tikzpicture}[node distance=2cm]
    \draw[gray] (0.5,2) node[left]{3}--(4,2);
    \draw[gray] (0.5,1) node[left]{2}--(4,1);
    \draw[gray] (0.5,0) node[left]{1}--(4,0);

    \draw[gray] (1,0) --(1,2.5) node[above]{4};
    \draw[gray] (2,0)--(2,2.5) node[above]{5};
    \draw[gray] (3,0)--(3,2.5) node[above]{6};
    \draw[gray] (4,0)--(4,2.5) node[above]{7};

    \node at (1.5,1.5) {$x_1$};
    \node at (2.5,1.5) {$x_3$};
    \node at (3.5,1.5) {$x_5$};
    \node at (1.5,0.5) {$x_2$};
    \node at (2.5,0.5) {$x_4$}; 
    \node at (3.5,0.5) {$x_6$};
    
    \draw[webarrow,thick] (0.5,1) -- (2,1);
    \draw[webarrow,thick] (2,1) -- (2,2);
    \draw[webarrow,thick] (2,2) -- (3,2);
    \draw[webarrow,thick] (3,2) -- (3,2.5);
    \draw[webarrow,thick] (0.5,0) -- (3,0);
    \draw[webarrow,thick] (3,0) -- (3,1);
    \draw[webarrow,thick] (3,1) -- (4,1);
    \draw[webarrow,thick] (4,1) -- (4,2.5);
  \end{tikzpicture}    
     \end{subfigure}
        \begin{subfigure}{0.3\textwidth}
 \begin{tikzpicture}[node distance=2cm]
    \draw[gray] (0.5,2) node[left]{3}--(4,2);
    \draw[gray] (0.5,1) node[left]{2}--(4,1);
    \draw[gray] (0.5,0) node[left]{1}--(4,0);

    \draw[gray] (1,0) --(1,2.5) node[above]{4};
    \draw[gray] (2,0)--(2,2.5) node[above]{5};
    \draw[gray] (3,0)--(3,2.5) node[above]{6};
    \draw[gray] (4,0)--(4,2.5) node[above]{7};

    \node at (1.5,1.5) {$x_1$};
    \node at (2.5,1.5) {$x_3$};
    \node at (3.5,1.5) {$x_5$};
    \node at (1.5,0.5) {$x_2$};
    \node at (2.5,0.5) {$x_4$}; 
    \node at (3.5,0.5) {$x_6$};
    
    \draw[webarrow,thick] (0.5,1) -- (2,1);
    \draw[webarrow,thick] (2,1) -- (2,2);
    \draw[webarrow,thick] (2,2) -- (3,2);
    \draw[webarrow,thick] (3,2) -- (3,2.5);
    \draw[webarrow,thick] (0.5,0)  -- (4,0);
    \draw[webarrow,thick] (4,0) -- (4,2.5);
  \end{tikzpicture}
       \end{subfigure}
       \begin{subfigure}{0.3\textwidth}
 \begin{tikzpicture}[node distance=2cm]
    \draw[gray] (0.5,2) node[left]{3}--(4,2);
    \draw[gray] (0.5,1) node[left]{2}--(4,1);
    \draw[gray] (0.5,0) node[left]{1}--(4,0);

    \draw[gray] (1,0) --(1,2.5) node[above]{4};
    \draw[gray] (2,0)--(2,2.5) node[above]{5};
    \draw[gray] (3,0)--(3,2.5) node[above]{6};
    \draw[gray] (4,0)--(4,2.5) node[above]{7};

    \node at (1.5,1.5) {$x_1$};
    \node at (2.5,1.5) {$x_3$};
    \node at (3.5,1.5) {$x_5$};
    \node at (1.5,0.5) {$x_2$};
    \node at (2.5,0.5) {$x_4$}; 
    \node at (3.5,0.5) {$x_6$};
    
    \draw[webarrow,thick] (0.5,1) -- (3,1);
    \draw[webarrow,thick] (3,1) -- (3,2.5);
    \draw[webarrow,thick] (0.5,0)  -- (4,0);
    \draw[webarrow,thick] (4,0) -- (4,2.5);
  \end{tikzpicture}
       \end{subfigure}
       
  \caption{Possible sets of non-intersecting paths from $\{1,2\}$ to $\{6,7\}$ describing the representation (\ref{webPlucker}) of the Pl\"ucker coordinate $p_{367}$ in ${\rm Gr}(3,7)$.}
   \label{webexpaths}
  \end{center}
\end{figure}
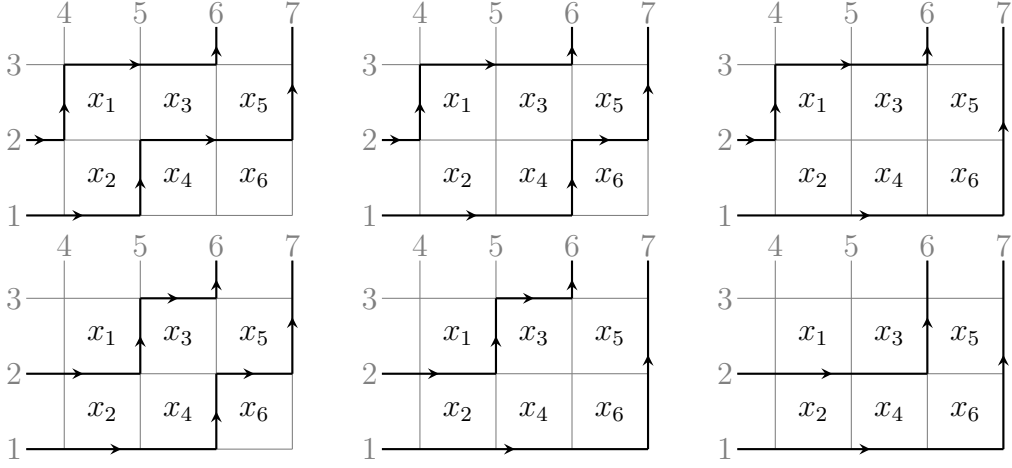

The procedure is best illustrated with an example: consider the Pl\"ucker coordinate $p_{367}$ in the case illustrated in Fig. \ref{webex}. We need to consider sets of non-intersecting paths from $\{1,2\}$ to $\{6,7\}$. We find the possible choices illustrated in Fig. \ref{webexpaths}.
The final result for the Pl\"ucker coordinate is therefore,
\begin{align}
p_{367} &=  x_1x_2x_3x_5 + x_1x_2x_3x_4x_5 + x_1x_2x_3x_4x_5x_6 \notag \\
&\quad + x_1^2x_2x_3x_4x_5  + x_1^2x_2x_3x_4x_5x_6 + x_1^2x_2x_3^2x_4x_5x_6\,.
\end{align}
To consider the tropical Grassmannian we tropicalise the resulting polynomial, replacing multiplication with addition and addition with minimum to obtain $w_K$.

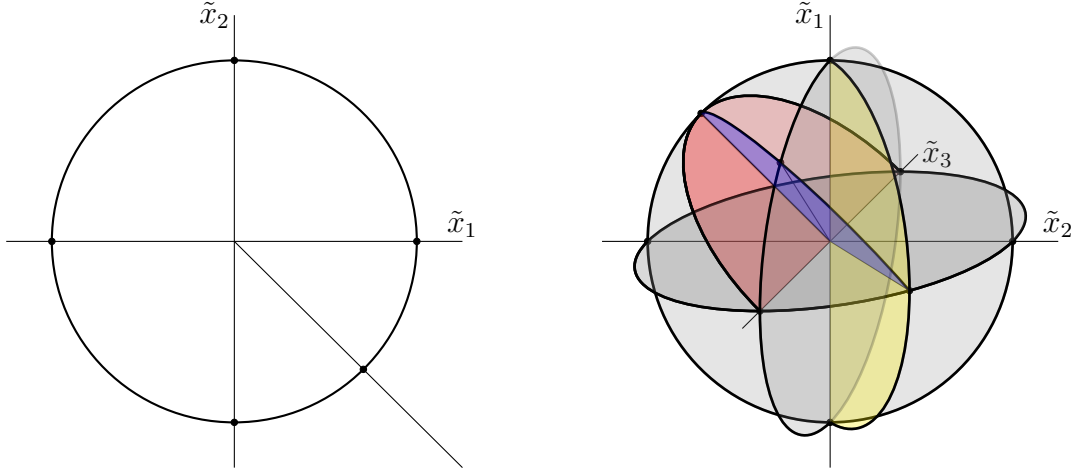
\begin{figure}[h]
  \centering
  \begin{subfigure}{0.4\textwidth}
    \begin{tikzpicture}[node distance=2cm, scale=0.6]
      \draw[opacity=0] (-5.2,-5,2) rectangle (5.2,5.2);
    \draw[] (-5,0) --(5,0) node[above]{$\tilde{x}_1$};
    \draw[] (0,-5) --(0,5) node[left]{$\tilde{x}_2$};
    \draw[] (0,0) --(5,-5) ;
    \draw[thick] (0,0) circle (4);
    \foreach \ang in {0, 90, 180, 270, 315}{
      \draw[fill=black] (\ang:4) circle (0.07);
    }
  \end{tikzpicture}
  \caption{The ${\rm Tr}^+(2,5)$ polyhedral fan}
  \label{fan25}
\end{subfigure}
\qquad \qquad
  \begin{subfigure}{0.4\textwidth} 
    \input{figures/fan26}
    \caption{The ${\rm Tr}^+(2,6)$ polyhedral fan}
    \label{fan26}
    \end{subfigure}
    \caption{The intersection of the ${\rm Tr}^+(2,n)$ fan with the unit sphere $S^{n-4}$ gives the dual of
      the ${\rm Gr}(2,n)$ associahedron. Notice the \TpGr{2,5} subfan on the
      $(\tilde x_1 , \tilde x_2)$ plane  of \TpGr{2,6}.}
  \label{G256fan}
\end{figure}

Following exactly the same logic for the simpler example of ${\rm Gr}(2,5)$ we obtain (as in \cite{2003math.....12297S})
\be
\begin{aligned}
p_{1i} &= p_{23} = 1\,, &&\qquad w_{1i} = w_{23} = 0\,, \notag \\
p_{24} &= 1+x_1, && \qquad w_{24} = {\rm min}(0,\tilde{x}_1)\,, \notag \\
p_{25} &= 1+x_1+x_1 x_2\,, && \qquad w_{25} = {\rm min}(0,\tilde{x}_1,\tilde{x}_1+\tilde{x}_2)\,, \notag \\
p_{34} &= x_1\,, &&\qquad w_{34} = \tilde{x}_1\,,\notag \\
p_{35} &= x_1 +x_1x_2\,, && \qquad w_{35} = {\rm min}(\tilde{x}_1,\tilde{x}_1+\tilde{x}_2)\,,\notag \\
p_{45} &= x_1 x_2\,, &&\qquad w_{45} = \tilde{x}_1+\tilde{x}_2\,.
\end{aligned}
\label{tropicalminorsG25}
\ee
The resulting tropical minors are piecewise linear functions in the space parametrised by $(\tilde{x}_1,\tilde{x}_2)$. Each such function defines tropical hypersurfaces in exactly the same way as before. Taking the union over the tropical hypersurfaces gives rise to a fan with five domains of linearity separated by five rays as illustrated in Fig.~\ref{fan25}.
We may label the five rays by 
\be
\{{\bf e}_1,{\bf e}_2,-{\bf e}_1,-{\bf e}_2,{\bf e}_1- {\bf e}_2\}
\label{Gr25SWrays}
\ee
where ${\bf e}_1$ and ${\bf e}_2$ are the two-component vectors,
\be
{\bf e}_1 = (1,0)\,,\qquad {\bf e}_2 = (0,1)\,.
\ee

More generally, the tropical minors in ${\rm Tr}^+(k,n)$ define a polyhedral fan in the $(k-1)(n-k-1)$-dimensional space of $\tilde{x}_i$ variables with many domains of linearity separated by walls of codimension one. The walls intersect in surfaces of codimension two and so on all the way down to individual rays of dimension one defined by the multiple intersection of (at least) $((k-1)(n-k-1)-1)$ walls. We illustrate the fan obtained in the case of ${\rm Tr}^+(2,6)$ in Fig. \ref{fan26}.

The five rays we have obtained correspond to the five positive rays among the set (\ref{G25rays}). We may verify this by evaluating the tropical minors $w_{ij}$ in (\ref{tropicalminorsG25}) on the five rays $\{{\bf e}_1,{\bf e}_2,-{\bf e}_1,-{\bf e}_2,{\bf e}_1-{\bf e}_2\}$. For example if we evaluate the ten-component vector of the $w_{ij}$ on ${\bf e}_1$ we obtain the vector
\be
{\bf e}_1 \mapsto {\rm ev}({\bf e}_1) = (0,0,0,0,0,0,0,1,1,1) \sim (1,0,0,0,0,0,0,0,0,0) = e_{12} \,.
\ee
where the equivalence corresponds to the lineality shift $w_{ij} \rightarrow w_{ij} +a_i +a_j$ with $(a_1,a_2,a_3,a_4,a_5) = \tfrac{1}{2}(1,1,-1,-1,-1)$.
Doing the same for each of the five rays in (\ref{Gr25SWrays}) we indeed obtain the ten-component vectors $\{e_{12},e_{45},e_{23},e_{15},e_{34}\}$, precisely the five positive rays in the list of ten solutions given in (\ref{G25rays}). The regions between the rays in Fig. \ref{fan25} then correspond to the edges between the positive rays in Fig. \ref{Petersen_graph}.

We may also recover the rays (\ref{Gr25SWrays}) from $\{e_{12},e_{45},e_{23},e_{15},e_{34}\}$ by tropically evaluating the coordinates $x_1$ and $x_2$ which are given by
\begin{align}
x_1 &= \frac{p_{12}p_{34}}{p_{14}p_{23}}\,,\quad  \longrightarrow \quad \tilde{x}_1 = w_{12} + w_{34} - w_{14} - w_{23}\,, \notag \\
x_2 &= \frac{p_{13}p_{45}}{p_{15}p_{34}}\,, \quad \longrightarrow \quad \tilde{x}_2 = w_{13} + w_{45} - w_{15} - w_{34}\,.
\end{align}
So for example the vector $e_{12}$ evaluates to $(1,0) = {\bf e}_1$ and the vector $e_{34}$ evaluates to $(1,-1) = {\bf e}_1-{\bf e}_2$.

Note that the rays (\ref{Gr25SWrays}) we have obtained from the tropical minors (\ref{tropicalminorsG25}) correspond to the dual vectors (\ref{G25dualvectors}) after dropping their first components. For example we have
\be
W_{24} = (c_{24} + c_{25},-1,0) \sim -{\bf e}_1\,.
\ee
The first component of the dual vector $W_{24}$ may be recovered by demanding for example
\be
Y \cdot W_{24} = y \cdot {\rm ev}(-e_1) = s_{23} = X_{24}\,,
\ee
where we recall $Y=(1,X_{13},X_{14})$ and $y=(s_{12},\ldots,s_{45})$.
Since the dual vectors are equivalent to the defining constraints of the kinematic associahedron, this gives us a way to recover the kinematic associahedron from the tropical minors.

The expressions of the web variables $x_i$ in terms of Pl\"ucker coordinates in fact identifies them with the cluster $\mathcal{X}$-variables of \cite{1021.16017,1054.17024} for the initial cluster of the ${\rm Gr}(2,5)$ cluster algebra. Indeed more generally the web variables are identified with the $\mathcal{X}$-coordinates of the initial cluster for any ${\rm Gr}(k,n)$. As we now outline, we can use the algebraic machinery of the cluster algebra to generate all the ray vectors describing the positive tropical Grassmannian ${\rm Tr}^+(k,n)$.

\section{The tropical Grassmannian and cluster algebras}

As mentioned above, we can identify cluster $\mathcal{X}$-coordinates
with web variables. As we shall see we can also identify the ray
vectors with cluster $\mathcal{A}$-coordinates. This allows us to
generalise the notion of mutation to these rays such that we can
generate all rays in the fan in a cluster algebraic way
\cite{2006math......2259F}. For a description of the relation between
cluster algebras and polyhedral fans, see also
\cite{readingtalk}. Before we demonstrate this it is useful to revisit
mutation for $\cA$-coordinates in Grassmannian cluster algebras.

A ${\rm Gr}(k,n)$ cluster is identified by its $m=(k-1)(n-k-1)$ unfrozen nodes, $n$ frozen nodes, and an $(m+n) \times (m+n)$ exchange matrix $B$ which encodes the connectivity of the nodes within the cluster. The first $m$ rows and columns correspond to the arrows between the unfrozen nodes.
Mutating an unfrozen node $k$ transforms $B$ to $B'$ given by

\begin{equation} \label{eq:bmutation}
b_{ij}' = \begin{cases}
	-b_{ij} & \text{if $i=k$ or $j=k$}. \\
	b_{ij} + [-b_{ik}]_+ b_{kj} + b_{ik} [b_{kj}]_+ & \text{otherwise}.
	\end{cases}
\end{equation}
where $[x]_+ = \max(x,0)$.
The mutated node also transforms, given by

\begin{equation}
a_k ' = \frac{1}{a_k} \prod^{m+n}_{i=1} a_{i}^{[b_{ik}]_+} + \prod^{m+n}_{i=1} a_{i}^{[-b_{ik}]_+} .
\end{equation}
Generalising mutations to rays requires additional information, namely an additional matrix $C$ (the coefficient matrix), its mutation given by\footnote{Note that we have modified slightly the mutation rule of the coefficient matrix of \cite{2006math......2259F} so that the ${\bf g}$ vectors defined by (\ref{ginitial},\ref{gmut}) match precisely the ray vectors for ${\rm Tr}^+(k,n)$ as defined in Sect. \ref{webs}.}
\begin{equation}
c_{ij}' = \begin{cases}
	-c_{ij} & \text{if $j=k$}. \\
	c_{ij} - [c_{ik}]_+ b_{kj} + c_{ik} [b_{kj}]_+ & \text{otherwise}.
	\end{cases}
\end{equation}
To each unfrozen $\mathcal{A}$-coordinate we associate a ray vector ${\bf g}$.
We start by constructing the initial cluster such that the $m$ unfrozen nodes are the $m$ basis vectors for $\mathbb{R}^m$ and $C$ is the identity
\begin{equation}
\label{ginitial}
\mathbf{g}_l = \mathbf{e}_l, \quad l = (1, \ldots, m), \quad C = \mathbb{I}_m.
\end{equation}
We then select a node $k$ to mutate on, following the mutation rule
\begin{align}
\mathbf{g}_l ' &= \mathbf{g}_l, \quad \text{for} \quad l \neq k \notag \\
\mathbf{g}_k ' &= -\mathbf{g}_k + \sum_{i=1}^{n} [-b_{ik}]_{+} \mathbf{g}_i + \sum_{j=1}^{n} [c_{jk}]_{+} \mathbf{b}_{j}^{0}
\label{gmut}
\end{align}
where $\mathbf{b}_{j}^{0} \text{, } j \in \{1, \ldots, m \} $ corresponds to the jth column of $B^0$, the exchange matrix for the initial cluster.
We can then repeat this process as many times as required to generate a vector for each unfrozen $\mathcal{A}$-coordinate. In the cases where the cluster algebra is of finite type (in this context the cases are ${\rm Gr}(2,n)$, ${\rm Gr}(3,6)$, ${\rm Gr}(3,7)$ and ${\rm Gr}(3,8)$) we obtain a finite cluster polytope by performing all mutations where each vertex is associated to a cluster. Each face of the polytope is associated to an unfrozen $\mathcal{A}$-coordinate $a$ and also by the above procedure a vector ${\bf g}$. 

The advantage of having the relation of the positive tropical fan to the cluster algebra is that it gives us a very easy algebraic way to generate the relevant ray vectors to describe the fan. Once we have the fan we can embed it into the original Pl\"ucker space using the tropical minors and compute its volume to obtain the generalised scattering amplitude.

\begin{figure}
	\centering
	{\footnotesize
  \begin{subfigure}{0.4\textwidth}
    \begin{tikzpicture}[node distance=10cm]
      \draw[line width=0, white] (-3.5,-3.5) rectangle (3.5,3.5);
		\node[draw, minimum size=4cm, regular polygon, 
    regular polygon sides=5,
    label=side 1:{$(0,1)$}, label=side 2:{$(-1,0)$}, label=side 3:{$(0,-1)$},
    label=side 4:{$(1,-1)$}, label=side 5:{$(1,0)$}] (B) {};
		\end{tikzpicture}
		\caption{The Tr$^+ (2,5)$ polytope labelled by rays.}
		\label{fig:tr25}
		\end{subfigure}
              }
              \quad\quad\quad
		{\footnotesize
		  \begin{subfigure}{0.4\textwidth}
                    \begin{tikzpicture}[scale=0.44]
                      
                            \draw[line width=0, white] ($({-3.5/0.44},{-3.5/0.44})+(5,0.25)$) rectangle ($({3.5/0.44},{3.5/0.44})+(5,0.25)$);
 \draw[join=bevel,thick,gray,dashed]  (5,-2.75) -- (5,-5.25);
 \draw[join=bevel,->,gray,dashed] (1.25,3.75) -- (2.5,2.5);
\node[gray] at (0.25,4.25){$(0,0,1)$};
\draw[join=bevel,->,gray,dashed] (8.75,3.75) -- (7.5,2.5);
\node[gray] at (9.75,4.25){$(1,-1,0)$};
\draw[join=bevel,->,gray,dashed] (8.75,-4.25) -- (7.5,-3);
\node[gray] at (9.75,-4.75){$(0,0,-1)$};
\draw[join=bevel,->,gray,dashed] (1.25,-4.25) -- (2.5,-3);
\node[gray] at (0.25,-4.75){$(-1,0,0)$};
 \draw[join=bevel,thick,gray,dashed] (0,0) -- (2.75,0.25) ;
  \draw[join=bevel,thick,gray,dashed]  (7.25,0.25) -- (10,0);
  \draw[join=bevel,thick,gray,dashed]  (2.75,0.25) -- (5,3.25) -- (7.25,0.25) -- (5,-2.75) -- cycle;
  \draw[join=bevel,thick,gray,dashed]  (5,3.25) -- (5,5);
  \draw[join=bevel,thick,fill=none] (0,0) -- (1,3) -- (2,-0.25) -- (1,-3.5)  -- cycle;
  \draw[join=bevel,thick,fill=none] (2,-0.25) -- (8,-0.25);
  \draw[join=bevel,thick,fill=none] (10,0) -- (9,3) -- (8,-0.25) -- (9,-3.5)  -- cycle;
  \draw[join=bevel,thick,fill=none] (1,3) -- (5,5) -- (9,3);
  \draw[join=bevel,thick,fill=none] (1,-3.5) -- (5,-5.25) -- (9,-3.5);
    \draw[thick,fill=none] (0,0) -- (1,3) -- (5,5) -- (9,3) -- (10,0) -- (9,-3.5) -- (5,-5.25) -- (1,-3.5) -- cycle;
\draw[join=bevel,->] (-0.75,-0.125) -- (1,-0.125);
\node at (-2,-0.125){(0,1,0)};
\draw[join=bevel,->] (10.75,-0.125) -- (9,-0.125);
\node at (12,-0.125){$(1,0,-1)$};
\draw[join=bevel,->] (5,5.75) arc (-200:-160:5.75);
\node at (5,6.5){$(1,0,0)$};
\draw[join=bevel,->] (5,-6) arc (-20:20:6.25);
\node at (5,-6.75){$(0,1,-1)$};
\node[gray] at (5,0.25){$(0,-1,0)$};
 \end{tikzpicture}
 \caption{The ${\rm Tr}^+(2,6)$ polytope with the faces labelled by rays.}
 \label{Stasheff}
	\end{subfigure}
	}
	\caption{The cluster polytopes pictured here are the dual polyhedra of those arising from the fans shown in Fig. \ref{G256fan}.}
\end{figure}
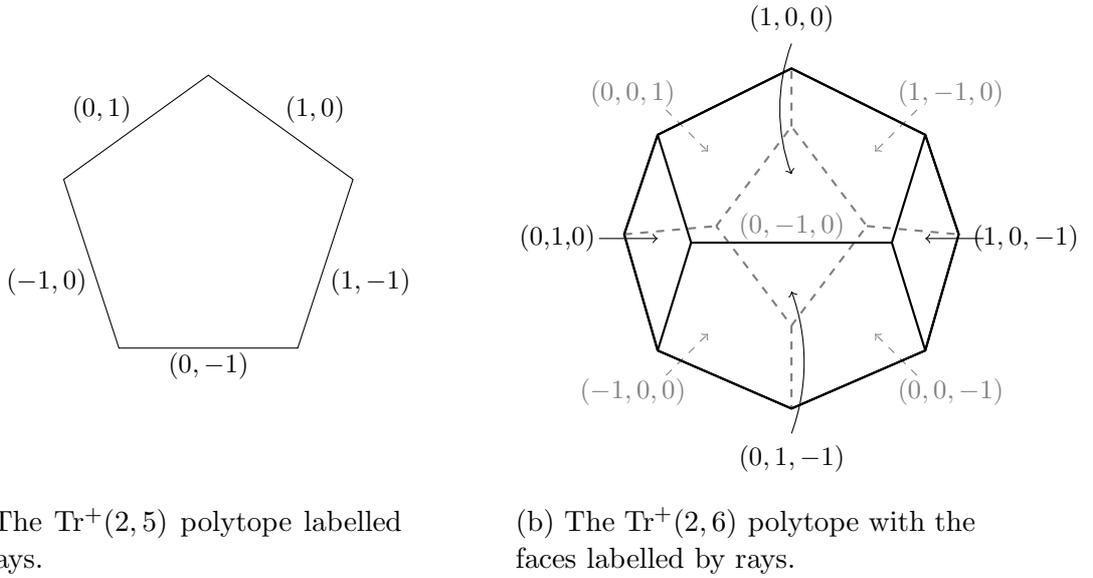

The resulting polytope in the simplest case is given in Figure \ref{fig:tr25}. 
It has five clusters connected in the shape of a pentagon. This pentagon is the dual of the pentagon obtained from intersecting the fan illustrated in Fig. \ref{fan25} with the unit circle; its edges are labelled with ray vectors (\ref{Gr25SWrays}).

In fact for ${\rm Gr}(2,n)$ the polytope obtained by intersecting the positive tropical fan with the unit sphere is always the dual polytope of the ${\rm Gr}(2,n)$ associahedron or Stasheff polytope. For example in Fig. \ref{Stasheff} we show the vectors associated to the faces of the $A_3$ associahedron. The dual polytope coincides with the intersection of the ${\rm Gr}(2,6)$ positive tropical fan with the unit sphere given in Fig. \ref{fan26}.

For the other finite cases the tropical positive fan gives polytopes that are closely related to the duals of the cluster polytopes as we now describe.

\section{\TpGr{3,6}}

Let us now consider the first case of the generalised biadjoint amplitudes which was addressed in \cite{Cachazo:2019ngv}. In analogy to the $\operatorname{Gr}{(2,n)}$ cases of the previous
section, the generalised amplitude for higher $k$ and $n$ can be interpreted as the volume of the 
computed by triangulating the relevant \TpGr{3,6} fan.

Following \cite{SpeyerSturmfels} we start by considering by the Pl\"ucker relations
of \Gr{3,6}, of which there are two kinds, three-term relations
and four-term relations,
  \begin{align}
    p_{123}p_{145} +p_{125}p_{134}  - p_{124}p_{135} &= 0, \dotsc \notag \\
    p_{123}p_{456} -p_{156} p_{234} + p_{146} p_{235} -  p_{145} p_{236}  & = 0, \dotsc
  \end{align}
While one can combinatorially generate many relations, only 35 of them
are linearly independent.

We then tropicalise these polynomials in Pl\"uckers to obtain
  \begin{align}
    \label{eq:tropPluckers36}
    &\min(w_{123}+w_{145},w_{125}+w_{134},w_{124}+w_{135}), \dotsc \notag \\
    &\min(w_{123}+w_{456},w_{156}+w_{234},w_{146}+w_{235},w_{145}+w_{236}),\dotsc\,.
  \end{align}
As before the tropical polynomials define
regions of linearity in the tropical Pl\"ucker space $\mathbb{R}^{20}$ separated by hypersurfaces defined as the
set of points at which the two smallest arguments of the $\min$
functions are equal. Consider for instance, the first tropical
polynomial in (\ref{eq:tropPluckers36}). It gives rise to a boundary
between two cones if one of the following is satisfied:
\begin{subequations}
\begin{align}
  w_{123}+w_{145} &= w_{125}+w_{134}\leq w_{124}+w_{135}\\
  \text{or}\qquad
  w_{123}+w_{145} &= w_{124}+w_{135}\leq w_{125}+w_{134}\\
  \text{or}\qquad
  w_{124}+w_{135} &=  w_{125}+w_{134}\leq w_{123}+w_{145}.
\end{align}
\end{subequations}
This polytope contains 65 vertices 
\cite{SpeyerSturmfels}. As above we denote the unit vectors in the $w_{ijk}$ directions by $e_{ijk}$. These vectors give 20 of the vertices. A further 15 are of the form
\be
f_{ijkl} = e_{ijk} + e_{jil} + e_{ikl} + e_{jkl}\,.
\ee
The remaining 30 are of the form (for $\{i,j,k,l,m,n\}$ distinct)
\be
g_{ij,kl,mn} = f_{ijkl} + e_{klm} + e_{kln}\,.
\ee

The part of the polytope that is relevant for a planar ordering is its
\emph{positive part} \TpGr{3,6}. In \cite{Cachazo:2019ngv} the positive vertices were determined by requiring compatibility with a planar ordering for the scattering amplitude. Here we identify the positive rays by
requiring that they satisfy the hypersurface conditions generated by monomials in the Pl\"ucker coordinates with opposite signs as we described in Sect. \ref{sect-tropical}.
This leaves us with 16 rays out of 65, coinciding precisely with the set of \cite{Cachazo:2019ngv}. They are $e_{123}$ and cyclic, $f_{1234}$ and cyclic and $g_{12,34,56}$, $g_{23,45,61}$, $g_{34,12,56}$ and $g_{45,23,61}$.

\begin{figure}
\begin{center}
  \begin{tikzpicture}[node distance=2cm]
    \draw (0.5,2) node[left]{3}--(3,2);
    \draw (0.5,1) node[left]{2}--(3,1);
    \draw (0.5,0) node[left]{1}--(3,0);

    \draw (1,0) --(1,2.5) node[above]{4};
    \draw (2,0)--(2,2.5) node[above]{5};
    \draw (3,0)--(3,2.5) node[above]{6};

    \node at (1.5,1.5) {$x_1$};
    \node at (2.5,1.5) {$x_3$};
    \node at (1.5,0.5) {$x_2$};
    \node at (2.5,0.5) {$x_4$};   
  \end{tikzpicture}
\caption{The web diagram for ${\rm Gr}(3,6)$.}
\label{Gr36web}
\end{center}
\end{figure}
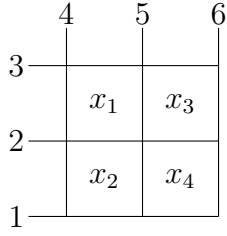

The \pGr{3,6} web diagram shown in Fig. \ref{Gr36web}
produces a matrix with following piecewise linear tropical minors \cite{2003math.....12297S,2015arXiv151102699B}, 
{\small
\begin{align}
w_{12i} &= w_{134} = w_{234} = 0 \,,\notag \\
w_{135} &= \min(0, \tilde{x}_1),\quad \notag \\
w_{136} &= \min(0, \tilde{x}_1, \tilde{x}_1 + \tilde{x}_3),\quad \notag \\
w_{145} &= \tilde{x}_1\,, \notag \\
w_{146} &= \min(\tilde{x}_1, \tilde{x}_1 + \tilde{x}_3) \,, \notag \\
w_{156} &= \tilde{x}_1 + \tilde{x}_3 \,,\notag \\
w_{235} &= \min(0, \tilde{x}_1, \tilde{x}_1 + \tilde{x}_2) \,, \notag \\
w_{236} &= 
  \min(0, \tilde{x}_1, \tilde{x}_1 + \tilde{x}_2, \tilde{x}_1 + \tilde{x}_3, \tilde{x}_1 + \tilde{x}_2 + \tilde{x}_3, \tilde{x}_1 + \tilde{x}_2 + \tilde{x}_3 + \tilde{x}_4) \,, \notag \\
w_{245} &= \min(\tilde{x}_1, \tilde{x}_1 + \tilde{x}_2) \,, \notag \\
w_{246} &= 
  \min(\tilde{x}_1, \tilde{x}_1 + \tilde{x}_2, \tilde{x}_1 + \tilde{x}_3, \tilde{x}_1 + \tilde{x}_2 + \tilde{x}_3, \tilde{x}_1 + \tilde{x}_2 + \tilde{x}_3 + \tilde{x}_4)\,, \notag \\
w_{256} &= \min(\tilde{x}_1 + \tilde{x}_3, \tilde{x}_1 + \tilde{x}_2 + \tilde{x}_3, \tilde{x}_1 + \tilde{x}_2 + \tilde{x}_3 + \tilde{x}_4) \,,\notag \\
w_{345} &= \tilde{x}_1 + \tilde{x}_2 \,, \notag \\
w_{346} &= \min(\tilde{x}_1 + \tilde{x}_2, \tilde{x}_1 + \tilde{x}_2 + \tilde{x}_3, \tilde{x}_1 + \tilde{x}_2 + \tilde{x}_3 + \tilde{x}_4) \,, \notag \\
w_{356} &= 
  \min(\tilde{x}_1 + \tilde{x}_2 + \tilde{x}_3, \tilde{x}_1 + \tilde{x}_2 + \tilde{x}_3 + \tilde{x}_4, 2 \tilde{x}_1 + \tilde{x}_2 + \tilde{x}_3 + \tilde{x}_4)\,, \notag \\
  w_{456} &= 2 \tilde{x}_1 + \tilde{x}_2 + \tilde{x}_3 + \tilde{x}_4\,.
  \label{eq:tropminors36} 
\end{align}
}
The regions of linearity of the tropical minors (\ref{eq:tropminors36}) define the fan for
\TpGr{3,6} and its intersection with the unit sphere $S^3$ is a
polytope with 16 vertices, 66 edges, 98 triangles and 48 three-dimensional facets.
The tropical $\mathcal{X}$-coordinates are given by
\begin{align}
\label{G36Xcoords}
\tilde{x}_1 &= w_{123} + w_{145} - w_{125} - w_{134}\,, \quad \tilde{x}_3 = w_{124} + w_{156} - w_{126} - w_{145} \,, \\
\tilde{x}_2 &= w_{124} + w_{345} - w_{145} - w_{234}\,, \quad \tilde{x}_4 = w_{134} + w_{125} + w_{456} - w_{124} - w_{156} - w_{345}\,. \notag
\end{align}
With the above relations (\ref{eq:tropminors36}) and (\ref{G36Xcoords}) we can go back and forth between the representation of the 16 positive vertices in terms of the $e_{ijk}$ and in terms of a four-component representation which we can also obtain from cluster mutations as we now describe.

\subsection{Triangulating \TpGr{3,6} with clusters}

Unlike in \Gr{2,5}, the \TpGr{3,6} fan contains facets that are not
simplicial. In particular, it contains 46 simplicial facets and two
bipyramids defined by five vertices. This is a common feature of $k>2$
(tropical) Grassmannians.

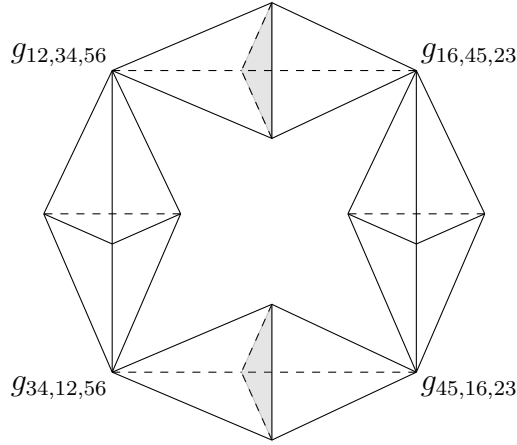
\begin{figure}[h]
  \centering
  \begin{tikzpicture}
    \draw (-2,2) node[above left] {$g_{12,34,56}$} -- ($(0,2)+(0.1,0.9)$);
    \draw (-2,2) -- ($(0,2)+(0.1,-0.9)$);
    \draw[dashed] (-2,2) -- ($(0,2)+(-0.3,0)$);

    \draw[line width=0, fill=gray, opacity=0.2] ($(0,2)+(0.1,0.9)$) -- ($(0,2)+(0.1,-0.9)$) -- ($(0,2)+(-0.3,0)$) -- cycle;
    \draw ($(0,2)+(0.1,0.9)$) -- ($(0,2)+(0.1,-0.9)$);
    \draw[dashed] ($(0,2)+(0.1,0.9)$) -- ($(0,2)+(-0.3,0)$);
    \draw[dashed] ($(0,2)+(0.1,-0.9)$) -- ($(0,2)+(-0.3,0)$);

    \draw[] (2,2) -- ($(0,2)+(0.1,0.9)$);
    \draw[] (2,2) -- ($(0,2)+(0.1,-0.9)$);
    \draw[dashed] (2,2) -- ($(0,2)+(-0.3,0)$);

    %%% 
    
    \draw (-2,-2) -- ($(0,-2)+(0.1,0.9)$);
    \draw (-2,-2) -- ($(0,-2)+(0.1,-0.9)$);
    \draw[dashed] (-2,-2) -- ($(0,-2)+(-0.3,0)$);

    \draw ($(0,-2)+(0.1,0.9)$) -- ($(0,-2)+(0.1,-0.9)$);
    \draw[dashed] ($(0,-2)+(0.1,0.9)$) -- ($(0,-2)+(-0.3,0)$);
    \draw[dashed] ($(0,-2)+(0.1,-0.9)$) -- ($(0,-2)+(-0.3,0)$);

    \draw[] (2,-2) node[below right] {$g_{45,16,23}$} -- ($(0,-2)+(0.1,0.9)$);
    \draw[] (2,-2) -- ($(0,-2)+(0.1,-0.9)$);
    \draw[dashed] (2,-2) -- ($(0,-2)+(-0.3,0)$);

    %%% 
    
    \draw (2,2) node[above right] {$g_{16,45,23}$} -- ($(2,0)+(0.9,0.1)$);
    \draw (2,2) -- ($(2,0)+(-0.9,0.1)$);
    \draw (2,2) -- ($(2,0)+(0,-0.3)$);

    \draw[line width=0, fill=gray, opacity=0.2] ($(0,-2)+(0.1,0.9)$) -- ($(0,-2)+(0.1,-0.9)$) -- ($(0,-2)+(-0.3,0)$) -- cycle;
    \draw[dashed] ($(2,0)+(0.9,0.1)$) -- ($(2,0)+(-0.9,0.1)$);
    \draw ($(2,0)+(-0.9,0.1)$) -- ($(2,0)+(0,-0.3)$);
    \draw ($(2,0)+(0,-0.3)$) --($(2,0)+(0.9,0.1)$);
    
    \draw (2,-2) -- ($(2,0)+(0.9,0.1)$);
    \draw (2,-2) -- ($(2,0)+(-0.9,0.1)$);
    \draw (2,-2) -- ($(2,0)+(0,-0.3)$);

    %%% 
    
    \draw (-2,2) -- ($(-2,0)+(0.9,0.1)$);
    \draw (-2,2) -- ($(-2,0)+(-0.9,0.1)$);
    \draw (-2,2) -- ($(-2,0)+(0,-0.3)$);

    \draw[dashed] ($(-2,0)+(0.9,0.1)$) -- ($(-2,0)+(-0.9,0.1)$);
    \draw ($(-2,0)+(-0.9,0.1)$) -- ($(-2,0)+(0,-0.3)$);
    \draw ($(-2,0)+(0,-0.3)$) --($(-2,0)+(0.9,0.1)$);
    
    \draw (-2,-2) node[below left] {$g_{34,12,56}$} -- ($(-2,0)+(0.9,0.1)$);
    \draw (-2,-2) -- ($(-2,0)+(-0.9,0.1)$);
    \draw (-2,-2) -- ($(-2,0)+(0,-0.3)$);

  \end{tikzpicture}
  \caption{The arrangement of bipyramids inside \TpGr{3,6}. The
    vertices represent the intersections of the rays $r_i$ with the
    unit sphere $S^4$. Two triangles are shaded to emphasize that they
    are actual 2-faces of the polytope.}
  \label{fig:bipyramids}
\end{figure}

To see this, first recall that $(k-1)(n-k+1)$ rays define a facet of
the fan if an arbitrary positive linear combination of them solves the
positive versions of inequalities derived from the Pl\"ucker
relations. 
In particular \TpGr{3,6} has 2 such facets with \emph{five} vertices
that form bipyramids. 
These non-simplicial bipyramids are arranged
inside the fan \TpGr{3,6} as sketched in Figure \ref{fig:bipyramids}.

The fan \TpGr{3,6} is closely related to the dual of the \pGr{3,6}
associahedron in that the latter provides a natural triangulation of
the former \cite{2015arXiv151102699B}. The vertices of the dual of the
associahedron correspond to cluster ${\mathcal A}$-coordinates. Two vertices are connected by an edge when
the corresponding pair of ${\cal A}$-coordinates appear together in
a cluster, i.e. are cluster-adjacent in the sense of
\cite{Drummond:2017ssj}. By definition, a pairwise connected
quadruplet of vertices of the dual \pGr{3,6} associahedron corresponds
to a cluster, which in turn can be identified as a simplex triangulating
\TpGr{3,6}.

We begin with the initial cluster
\begin{equation}
\begin{tikzpicture}%
    [
    unfrozen/.style={},
    frozen/.style={inner sep=1.5mm,outer sep=0mm,yshift=0},
    node distance = 0.4cm
    ]
    \node[frozen]  (t1)  at (1,3) {$(124)$};
    \node[frozen, right = of t1]        (t2)  {$(125)$};
  %  \node[frozen, right = of t2]        (t3)  {$(126)$};
  %  \node[frozen, right = of t3]        (t4)  {$(127)$};
    %
    \node[frozen, below = of t1]        (b1)  {$(134)$};
    \node[frozen, below = of t2]        (b2)  {$(145)$};
  %  \node[frozen, below = of t3]        (b3)  {$(156)$};
  %  \node[frozen, below = of t4]        (b4)  {$(167)$};
    
    \draw[->] (t1) -- (t2); 
    %\draw[->] (t2) -- (t3); \draw[->] (t3) -- (t4);
    \draw[->] (b1) -- (b2); 
    %\draw[->] (b2) -- (b3); \draw[->] (b3) -- (b4);
    \draw[->] (t1) -- (b1); \draw[->] (t2) -- (b2); 
    %\draw[->] (t3) -- (b3); \draw[->] (t4) -- (b4);
    \draw[->] (b2) -- (t1); 
    %\draw[->] (b3) -- (t2); \draw[->] (b4) -- (t3); 
  \end{tikzpicture}
  \notag
\end{equation}
and associate its unfrozen
$\mathcal{A}$-coordinates with the unit vectors
${\mathbf e}_1,\dotsc, {\mathbf e}_4$ in the tropical $\tilde{x}_i$ coordinates,
\begin{align}
  (124)\leftrightarrow {\bf e}_1   &=(1,0,0,0),& (125)&\leftrightarrow \makebox[0.6cm][c]{${\bf e}_2$}=(0,1,0,0),\\
  (134)\leftrightarrow {\bf e}_3   &=(0,0,1,0),& (145)&\leftrightarrow \makebox[0.6cm][c]{${\bf e}_4$}  =(0,0,0,1)\,.
\end{align}
Performing all possible mutations generates the full set of 16 ray vectors which arise from 50 distinct clusters.

Among the 16 rays we find the following five,
\begin{align}
   & \makebox[0.6cm][c]{${\bf e}_{3}$} &\mapsto &\quad
   (\makebox[0.42cm]{0}\makebox[0.42cm]{0}\makebox[0.42cm]{0}\makebox[0.42cm]{0}\makebox[0.42cm]{0}\makebox[0.42cm]{0}\makebox[0.42cm]{0}\makebox[0.42cm]{0}\makebox[0.42cm]{0}\makebox[0.42cm]{1}\makebox[0.42cm]{0}\makebox[0.42cm]{0}\makebox[0.42cm]{0}\makebox[0.42cm]{0}\makebox[0.42cm]{0}\makebox[0.42cm]{1}\makebox[0.42cm]{0}\makebox[0.42cm]{0}\makebox[0.42cm]{1}\makebox[0.42cm]{1}) \sim f_{1234}\nonumber\\
    &\makebox[0.6cm][c]{$-{\bf e}_{1}$} &\mapsto &\quad
   (\makebox[0.42cm]{0}\makebox[0.42cm]{0}\makebox[0.42cm]{0}\makebox[0.42cm]{0}\makebox[0.42cm]{0}\makebox[0.42cm]{-1}\makebox[0.42cm]{-1}\makebox[0.42cm]{-1}\makebox[0.42cm]{-1}\makebox[0.42cm]{-1}\makebox[0.42cm]{0}\makebox[0.42cm]{-1}\makebox[0.42cm]{-1}\makebox[0.42cm]{-1}\makebox[0.42cm]{-1}\makebox[0.42cm]{-1}\makebox[0.42cm]{-1}\makebox[0.42cm]{-1}\makebox[0.42cm]{-2}\makebox[0.42cm]{-2}) \sim f_{1256} \nonumber\\
     &\makebox[0.6cm][c]{${\bf e}_{2}$} &\mapsto &\quad
    (\makebox[0.42cm]{0}\makebox[0.42cm]{0}\makebox[0.42cm]{0}\makebox[0.42cm]{0}\makebox[0.42cm]{0}\makebox[0.42cm]{0}\makebox[0.42cm]{0}\makebox[0.42cm]{0}\makebox[0.42cm]{0}\makebox[0.42cm]{0}\makebox[0.42cm]{0}\makebox[0.42cm]{0}\makebox[0.42cm]{0}\makebox[0.42cm]{0}\makebox[0.42cm]{0}\makebox[0.42cm]{0}\makebox[0.42cm]{1}\makebox[0.42cm]{1}\makebox[0.42cm]{1}\makebox[0.42cm]{1}) \sim f_{3456} \nonumber\\
    &\makebox[0.6cm][c]{${\bf e}_{2}+{\bf e}_{3}-{\bf e}_{4}$} &\mapsto &\quad
    (\makebox[0.42cm]{0}\makebox[0.42cm]{0}\makebox[0.42cm]{0}\makebox[0.42cm]{0}\makebox[0.42cm]{0}\makebox[0.42cm]{0}\makebox[0.42cm]{0}\makebox[0.42cm]{0}\makebox[0.42cm]{0}\makebox[0.42cm]{1}\makebox[0.42cm]{0}\makebox[0.42cm]{0}\makebox[0.42cm]{0}\makebox[0.42cm]{0}\makebox[0.42cm]{0}\makebox[0.42cm]{1}\makebox[0.42cm]{1}\makebox[0.42cm]{1}\makebox[0.42cm]{1}\makebox[0.42cm]{1})
     \sim g_{12,34,56} \nonumber\\
    &\makebox[0.6cm][c]{${\bf e}_{4}-{\bf e}_{1}$} &\mapsto &\quad
    (\makebox[0.42cm]{0}\makebox[0.42cm]{0}\makebox[0.42cm]{0}\makebox[0.42cm]{0}\makebox[0.42cm]{0}\makebox[0.42cm]{-1}\makebox[0.42cm]{-1}\makebox[0.42cm]{-1}\makebox[0.42cm]{-1}\makebox[0.42cm]{-1}\makebox[0.42cm]{0}\makebox[0.42cm]{-1}\makebox[0.42cm]{-1}\makebox[0.42cm]{-1}\makebox[0.42cm]{-1}\makebox[0.42cm]{-1}\makebox[0.42cm]{-1}\makebox[0.42cm]{-1}\makebox[0.42cm]{-1}\makebox[0.42cm]{-1})     \sim g_{34,12,56}\notag
\end{align}
where we have also given their evaluations through the tropical minors (\ref{eq:tropminors36}) and the corresponding positive solutions given above. The fact that these five vertices form a single bipyramid rather than two tetrahedral facets can be seen from the linear relation,
\be
f_{1234} + f_{1256} + f_{3456} = g_{12,34,56} + g_{34,12,56}\,.
\ee

Note that the cluster algebra provides a canonical way of determining a triangulation. In
particular the bipyramid formed by the five rays described above is triangulated by two clusters whose vertices are given by $\{f_{1234},f_{1256},f_{3456},g_{12,34,56}\}$ and $\{f_{1234},f_{1256},f_{3456},g_{34,12,56}\}$.

Equipped with the cluster triangulation, we can express the scattering
amplitude as a sum over clusters,

\begin{equation}
  m_3^6(\mathbb{I}|\mathbb{I})
  =
  \sum_{c\, \in\, \begin{minipage}{1.2cm}\begin{spacing}{0.1}\tiny clusters of \pGr{3,6}\end{spacing}\end{minipage}}
  \,\,\,\,
  \prod_{a\,\in\,\begin{minipage}{1.6cm}\begin{spacing}{0.1}\centering \tiny $\mathcal A$-coords  of $c$\end{spacing}\end{minipage} }\,
  \frac{1}
  {y \cdot {\rm ev}(r_a)}\,,
\end{equation}
where as before $y=(s_{123},\ldots,s_{456})$ is the lexicographically ordered vector of Mandelstam invariants, $r_a$ is the representation of the $\mathcal A$-coordinate $a$
as a ray in $\tilde{x}$ coordinates and ${\rm ev}$ means the evaluation using the 
tropical minors in (\ref{eq:tropminors36}).

Using this identification, we can read off the two terms in the
amplitude directly from the two clusters as\footnote{Here we use the notation $t_{ijkl} = s_{ijk} + s_{ijl} + s_{ikl} + s_{jkl}$ and $R_{ij,kl,mn} = t_{ijkl} + s_{klm} + s_{kln}$.}

\begin{equation}
  \label{eq:twoterm}
  \frac{1}{t_{1234} t_{1256}t_{3456}}\,
  \biggl[
  \frac{1}{  R_{12,34,56}} +  \frac{1}{R_{34,12,56}} 
  \biggr]\, .
\end{equation}
Note that using the identity between kinematic invariants
\begin{equation}
  R_{12,34,56}+R_{34,12,56}
  =
  t_{1234}+t_{3456}+t_{5612}
\end{equation}
we can write these two terms as 
\begin{equation}
  (\ref{eq:twoterm})
  =
  \frac{1}{R_{12,34,56}R_{34,12,56}}
  \biggl[
  \frac{1}{t_{1256}t_{3456}}
  +\frac{1}{t_{1234} t_{3456}}
  +\frac{1}{t_{1234} t_{1256}}\biggr]
\end{equation}
which was noted in \cite{Cachazo:2019ngv} to correspond to a different
triangulation of the bipyramid. However the cluster algebra prefers a
particular one of these triangulations.

\section{${\rm Tr}(3,7)$: the amplitude from $E_6$ clusters}
In this section we explicitly demonstrate how the triangulation of the
fan associated to the positive tropical Grassmannian
\TpGr{3,7} can be worked out from the \Gr{3,7} cluster algebra.

As in the previous section, one can either compute $F_{3,7}$ from the
web $\operatorname{Web}_{3,7}$ or run the cluster-algebra machinery to
obtain the generalised amplitude without even referring to
\TGr{3,7}. Nevertheless let us first describe \TpGr{3,7} starting from
\TGr{3,7} and elaborate on a situation that is not encountered in
Grassmannians of lower dimension.

The tropical Grassmannian \TGr{3,7} has 721 rays which come in six
types,\footnote{This form was also given in \cite{Cachazo:2019apa}.}
\begin{subequations}
  \begin{align}
      \label{eq:bee1}
      b_{1,1234567} & = e_{123}, \\
    b_{2,1234567} & = e_{123}+e_{124}+e_{134}+e_{234},\\
    b_{3,1234567} & = e_{123}+e_{124}+e_{125}+e_{126}+e_{127}, \\
    b_{4,1234567} & = e_{123}+e_{124}+e_{125}+e_{126}+e_{127}+e_{134}+e_{234}, \\
    b_{5,1234567} & = e_{123}+e_{124}+e_{125}+e_{126}+e_{127}+e_{134}+e_{156}+e_{234}+e_{256}\\
    b_{6,1234567} & = b_{3,1234567}+b_{3,3456712}+b_{3_,6712345}
\,,
    %
    % 
  % b_{1,1,2,3,4,5,6,7}
  % &=
  %   s_{123} \label{eq:bee1}\\
  % b_{2,123456}
  % &=
  %   \sum_{1\leq i,j,k\leq 4}s_{ijk}\\
  % b_{3,123456}
  % &=\sum_{1\leq i,j,k\leq 5}s_{ijk}\\
  % b_{4,123456}
  % &=\sum_{i=3}^7s_{12i}+s_{134}+s_{234}\\
  % b_{5,123456}
  % &=b_{4,123456}+s_{156}+s_{256}\\
  % b_{6,123456}
  % &=b_{3,123456}+b_{3,345671}+b_{3,671234}
    \label{eq:bee6}
\end{align}
\end{subequations}
where as before the lexicographically-ordered $e_{ijk}$ are identified with unit
vectors in $\mathbb{R}^{7 \choose 3}$. Other rays are obtained by the
permutations of those that are written out above. For the types
$b_1,\dotsc, b_5$ and $b_6$, the permutations generate the six
symmetry classes of respective sizes 35, 35, 21, 210, 315 and
105. These rays have also been tabulated in \cite{planes37} with their
explicit Pl\"ucker coordinates.
Henceforth we will drop the labels in $b_{i,1234567}$ and just write
$b_i$ unless the order of the indices is not canonical.

To compute positive Grassmannian \TpGr{3,7}, we select out of the 721
rays above those which solve the positive versions of tropicalised
Pl\"ucker relations. One
finds that 49 of them satisfy such relations. This seems incompatible
with the fact that the cluster algebra has 42 distinct unfrozen $\mathcal{A}$-coordinates.

The resolution to this discrepancy is that seven positive rays of the
type $b_6$ are linear combinations of three mutually-connected rays of
type $b_3$, any positive linear combination of which is a solution. In
other words, $b_{6,1234567}$ is in the middle of a triangular 2-face of
\TpGr{3,7} and is not necessary to define a cone of the fan.

As explained by Speyer and Williams \cite{2003math.....12297S}, the
${\rm Tr}^+(3,7)$ fan has 693 facets. While 595 of these facets are
simplicial, there are also 63 facets with 7 vertices, 28 with 8
vertices and 7 with 9 vertices. These non-simplicial facets are
the $\operatorname{Gr}(3,7)$ analogues of the bipyramids of ${\rm Tr}^+(3,6)$.

We again resort to the relevant cluster algebra $E_6$ to obtain a
triangulation on which we evaluate the amplitude. The $E_6$ cluster
algebra has 833 clusters that give the vertices of the associahedron. These
833 clusters make up the simplexes of the triangulation each of which
contain six vertices. 

If we employ the duality between \Gr{3,7} and \Gr{4,7} and work in
terms of the latter, we can relate the positive vertices above to the established notation for $\mathcal{A}$-coordinates in the literature on $\mathcal{N}=4$ amplitudes \cite{Drummond:2014ffa,Drummond:2017ssj}. The different types of rays classified in
(\ref{eq:bee1})-(\ref{eq:bee6}) nicely match the conventional cluster
$\mathcal{A}$-coordinates:
\begin{align}
  \label{eq:acoordsasrays47}
  a_{11} &\leftrightarrow b_{2,7123456}& a_{41} &\leftrightarrow b_{4,7156234}\nonumber\\
  a_{21} &\leftrightarrow b_{1,7123456}& a_{51} &\leftrightarrow b_{4,2345671}\nonumber\\
  a_{31} &\leftrightarrow b_{3,5671234}& a_{61} &\leftrightarrow b_{5,1234675}\,,
\end{align}
where the rest of the correspondence can be worked out by cyclic
rotations of the second indices of the $a_{ij}$ and the arguments of
the $b_i$. With this correspondence, we find that the $E_6$ initial cluster
\begin{equation}
\begin{tikzpicture}%
    [
    unfrozen/.style={},
    frozen/.style={inner sep=1.5mm,outer sep=0mm,yshift=0},
    node distance = 0.4cm
    ]
    \node[frozen]  (t1)  at (1,4) {$a_{24}$};
    \node[frozen, right = of t1]        (t2)  {$a_{37}$};
    \node[frozen, below = of t1]        (m1)  {$a_{13}$};
    \node[frozen, below = of t2]        (m2)  {$a_{17}$};
    \node[frozen, below = of m1]        (b1)  {$a_{32}$};
    \node[frozen, below = of m2]        (b2)  {$a_{27}$};
    
    \draw[->] (t1) -- (t2); \draw[->] (t1) -- (m1); \draw[->] (t2) -- (m2);
    \draw[->] (m1) -- (m2); \draw[->] (m1) -- (b1); \draw[->] (m2) -- (t1); \draw[->] (m2) -- (b2) ;
    \draw[->] (b1) -- (b2); \draw[->] (b2) -- (m1);      

  \end{tikzpicture}
\end{equation}
produces the following term in the amplitude 
\begin{equation}
  \frac{1}
  {
(y \cdot b_{1,1234567})(y \cdot b_{3,1234567})(y \cdot b_{2,1234567})(y \cdot b_{2,4567123})(y \cdot b_{3,6712345})(y \cdot b_{1,5671234})
}\,,
\end{equation}
with $y=(s_{123},\ldots, s_{567})$.
We then mutate these rays according to (\ref{gmut}) iteratively until
we cover all 833 clusters of the $E_6$ polytope. Recovering the
corresponding kinematic invariants using (\ref{eq:acoordsasrays47}),
we can construct the \Gr{3,7} amplitude as the volume of the positive
tropical Grassmannian. An expression for this amplitude is provided in
the ancillary file \verb+Gr37amp.m+ .

\section{\Gr{3,8}: redundant triangulations}
In this section we will run the same construction in \Gr{3,8} to
provide a conjecture for the canonically-ordered part of the
generalised biadjoint amplitude that one would obtain by solving the
scattering equations for this Grassmannian.

We start with the initial cluster of \Gr{3,8}
\begin{equation}
\begin{tikzpicture}%
    [
    unfrozen/.style={},
    frozen/.style={inner sep=1.5mm,outer sep=0mm,yshift=0},
    node distance = 0.4cm
    ]
    \node[frozen]  (t1)  at (1,3) {$(124)$};
    \node[frozen, right = of t1]        (t2)  {$(125)$};
    \node[frozen, right = of t2]        (t3)  {$(126)$};
    \node[frozen, right = of t3]        (t4)  {$(127)$};
    \node[frozen, below = of t1]        (b1)  {$(134)$};
    \node[frozen, below = of t2]        (b2)  {$(145)$};
    \node[frozen, below = of t3]        (b3)  {$(156)$};
    \node[frozen, below = of t4]        (b4)  {$(167)$};
    
    \draw[->] (t1) -- (t2); \draw[->] (t2) -- (t3); \draw[->] (t3) -- (t4);
    \draw[->] (b1) -- (b2); \draw[->] (b2) -- (b3); \draw[->] (b3) -- (b4);
    \draw[->] (t1) -- (b1); \draw[->] (t2) -- (b2); \draw[->] (t3) -- (b3); \draw[->] (t4) -- (b4);
    \draw[->] (b2) -- (t1); \draw[->] (b3) -- (t2); \draw[->] (b4) -- (t3); 
  \end{tikzpicture}
\end{equation}
and identify its $\mathcal{A}$-coordinates with the
rays $(124) \leftrightarrow \mathbf{e_1}$,
\mbox{$(134), \leftrightarrow \mathbf{e_2}$},
\mbox{$(125) \leftrightarrow \mathbf{e_3}$}, \ldots,
$(167) \leftrightarrow \mathbf{e_8}$ in $\mathbb{R}^8$. Using the map
explained in Sect. \ref{Gr25SWrays} we recover the Pl\"ucker
coordinates for the Speyer-Williams rays
$\mathbf{e}_1,\dotsc, \mathbf{e}_8$ and deduce that the initial
cluster produces the following for the first term in the amplitude
\begin{align}
&1/\bigl((y \cdot b_{1,1 2 3 4 5 6 7 8}) (y \cdot b_{3,1 2 3 4 5 6 7 8})
 (y \cdot b_{2,1 2 3 4 5 6 7 8})(y \cdot b_{5,6 7 5 4 8 1 2 3}) \notag \\
&\quad  \times (y \cdot b_{5,3 4 2 1 5 6 7 8})(y \cdot b_{2,5 6 7 8 1 2 3 4})(y \cdot b_{3,7 8 1 2 3 4 5 6}) (y \cdot b_{1,6 7 8 1 2 3 4 5})\bigr)\, ,
\end{align}
where the $b$ vertices are given below in (\ref{Gr38bs}) and as before $y=(s_{123},\ldots,s_{678})$.

We then generate all 25,080 clusters using the mutation rules of
\cite{2006math......2259F} which we have adapted in equation
(\ref{gmut}). These clusters contain 128 distinct vectors in
$\mathbb{R}^8$, identified with the 128 $\mathcal{A}$-coordinates of
\Gr{3,8}. As usual, the Pl\"ucker coordinates of these vectors
provides us the factors in the denominator of every term in the
amplitude. We provide all 25080 terms in the ancillary file
\verb+Gr38amp.m+ .

Let us comment further on the correspondence between the ${\rm Tr}^+(3,8)$ fan and the
\Gr{3,8} cluster algebra. We find that, out of the 128 vectors
generated by the cluster algebra, only 120 are rays of the
corresponding fan. The extra 8 vectors have the form
\begin{equation}
b_{\mathrm e} = b_{8,12345678}+b_{8,78564123}
\end{equation}
and cyclic rotations thereof. These too
are positive vectors but being linear
combinations of two genuine rays
they lie on an edge of the fan. In other words, they
separate only 7 regions of piecewise linearity for the tropical minors instead of 8. This can
be interpreted as the \Gr{3,7} cluster algebra producing
\emph{redundant triangulation} of the fan which decomposes already
simplicial facets into even smaller simplexes.

We can compare the Pl\"ucker coordinates of the vectors we obtain to
the rays of another object called the Dressian \Dr{3,8}, studied in
\cite{Dressian38}. \Dr{3,8} is a non-simplicial fan that consists of
15470 rays which split into 12 symmetry classes of size (56, 70, 28,
420, 56, 1260, 420, 560, 1680, 840, 5040, 5040). These define facets
in groups of sizes ranging from 8 to 12. While all rays of \Dr{k,n}
are expected to be rays of \Gr{k,n}, the converse is not true. Indeed
the Dressian \Dr{3,8} does not capture the rays $b_8$ which give rise
to ``superfluous'' triangulations.

The rays of \Dr{3,8}, positive and non-positive, are explicitly given as:
\begin{equation}
\begin{aligned}
  b_1 & = e_{123}, \\
  b_2 & = e_{123}+e_{124}+e_{134}+e_{234},\\
  b_3 & = e_{123}+e_{124}+e_{125}+e_{126}+e_{127}+e_{128}, \\
  b_4 & = e_{123}+e_{124}+e_{125}+e_{126}+e_{127}+e_{128}+e_{134}+e_{234}, \\
  b_5 & = e_{123}+e_{124}+e_{125}+e_{134}+e_{135}+e_{145}+e_{234}+e_{235}+e_{245}+e_{345}, \\
  b_6 & = e_{123}+e_{124}+e_{125}+e_{126}+e_{127}+e_{128}+e_{136}+e_{145}+e_{236}+e_{245}, \\
  b_7 & = e_{123}+e_{124}+e_{125}+e_{126}+e_{127}+e_{128}+e_{138}+e_{147}+e_{156}+e_{238}\\
  &~ +e_{247}+e_{256}, \\
  b_8 & = e_{123}+e_{124}+e_{125}+e_{126}+e_{127}+e_{128}+e_{134}+e_{135}+e_{145}+e_{234}\\
  &~ +e_{235}+e_{245}+e_{345}, \\
  b_9 & = e_{123}+e_{124}+e_{125}+e_{126}+e_{127}+e_{128}+e_{134}+e_{135}+e_{145}+e_{167}\\
  &~ +e_{234}+e_{235}+e_{245}+e_{267}+e_{345}, \\
  b_{10} & = e_{123}+e_{124}+e_{125}+e_{134}+e_{135}+e_{145}+e_{146}+e_{147}+e_{148}+e_{234}\\
  &~+e_{235}+e_{236}+e_{237}+e_{238}+e_{245}+e_{345}, \\
  b_{11} & = e_{123}+e_{124}+e_{125}+e_{126}+e_{127}+e_{128}+e_{134}+e_{137}+e_{147}+e_{156}\\
  &~+e_{234}+e_{237}+e_{247}+e_{256}+e_{345}+e_{346}+e_{347}+e_{348}, \\
  b_{12} & = e_{123}+e_{124}+e_{125}+e_{126}+e_{127}+e_{128}+e_{134}+e_{138}+e_{148}+e_{157}\\
  &~+e_{234}+e_{238}+e_{248}+e_{257}+e_{345}+e_{346}+e_{347}+e_{348}+e_{356}+e_{456}.
\label{Gr38bs}
\end{aligned}
\end{equation}
Out of these, the 120 vectors defined by
\begin{equation}
\begin{aligned}
\{&b_{1,12345678},b_{2,12345678},b_{3,12345678},b_{4,12345678},b_{4,23184567},b_{5,23184567},\\
&b_{6,12378456},b_{8,12345678},b_{8,34128567},b_{9,12345786},b_{9,23178456},b_{10,13482567},\\
&b_{11,34185627},b_{11,12457836},b_{12,12457683} \} 
\end{aligned}
\end{equation}
and their cyclic copies
lie in the positive region in the sense that they satisfy the positive
version of the inequalities (\ref{tropical_conditions}). These
vectors are in one-to-one correspondence with the 120 non-redundant
rays generated by the cluster algebra.

Note that the redundant vectors $b_{\mathrm e}$ that we encountered in
\Gr{3,8} are of different nature to the $b_6$ of \Gr{3,7}. While both
types of vectors are not rays of the relevant fan, unlike the $b_\mathrm e$, the $b_6$ are not generated by the
cluster algebra.

\section{Conclusions and outlook to \Gr{4,8}}
In this paper we have utilised cluster algebra technology to construct
tree-level biadjoint amplitudes on ${\rm Gr}(3,n)$ for $n=6,7,8$. These
amplitudes arise from scattering equations on the corresponding
Grassmannians \cite{Cachazo:2019ngv, Cachazo:2019apa} and the
relevance of cluster algebras for these amplitudes arises from the
interpretation of these amplitudes as volumes of certain geometric
objects. In the cases we studied in this paper these objects are
polyhedra in $(k-1)(n-k-1)-1$ dimensions, where $k=3$.

Cluster algebras provide a natural triangulation of the polyhedra
whose volumes correspond to the scattering amplitudes. Therefore we
were able to employ mutation rules to ``bootstrap'' the amplitude
starting from a single term only. In particular we provided a
prescription for the volume of the simplex that corresponds to the
initial cluster and obtained the volumes of the remaining simplexes
through consecutive cluster mutations.

Each of the cases we considered has new features that provide
important lessons. In $n=6$ we saw that the clusters triangulate the
bipyramids of \TpGr{3,6} into two simplexes. In $n=7$ we identified that
positive rays that define cones of \TpGr{3,7} are not rays of the
${\rm Tr}^+(3,7)$ fan and are also not detected by the cluster algebra. When we studied the
$n=8$ case, we found that the cluster algebra generates redundant
triangulations of the ${\rm Tr}^+(3,8)$ fan. 

Having studied the fans corresponding to various Grassmannians, a
natural direction to take is to attempt to construct the fan for ${\rm Tr}^+(4,8)$, which corresponds to the positive part
of \Gr{4,8}. The cluster algebra of the latter is expected to
capture the rational symbol letters of 8-particle amplitudes in
$\mathcal N=4$ but the fact that the \Gr{4,8} cluster algebra is infinite has been a forbidding obstacle in utilising cluster algebras in the computations of these
amplitudes.

We find that restricting the mutations to clusters that contain only
rays obeying the full number of intersection conditions for
${\rm Tr}^+(4,8)$ closes on a finite number of 169,192 clusters. The
corresponding $\mathcal A$-coordinates in these clusters provides us
with a finite alphabet of 356 rational letters closed under cyclic rotations
of the twistors. In particular, this alphabet contains the rational
letters reported in \cite{Prlina:2017tvx}. It would be interesting to
check if these letters are in correspondence with the faces of the
polytope found by Arkani-Hamed, Lam and Spradlin, as reported in
\cite{VolovichXMAS18}.

\section*{Acknowledgments}
All authors are supported by ERC grant 648630 IQFT.

\Urlmuskip=0mu plus 1mu\relax
\def\UrlBreaks{\do\/\do-}
\bibliographystyle{JHEP}
\bibliography{biblio}

\end{document}

%% file: figures/fan26.tex
\begin{tikzpicture}[scale=0.6]
      \draw[opacity=0] (-5.2,-5,2) rectangle (5.2,5.2);

% \pgfsetyvec{\pgfpoint{1cm}{0}}
% \pgfsetxvec{\pgfpoint{0cm}{1cm}}
% \pgfsetzvec{\pgfpoint{0}{1cm}}
  
  \draw (0,0,0) -- (5,0,0) node[above] {$\tilde x_2$};
  \draw (0,0,0) -- (-5,0,0);
  \draw (0,0,0) -- (0,5,0)node[left] {$\tilde x_1$};
  \draw (0,0,0) -- (0,-5,0);
  \draw (0,0,0) -- (0,0,5);
  \draw (0,0,0) -- (0,0,-5)node[right] {$\tilde x_3$};
  \draw (0,0,0) -- ({4/sqrt(2)},0,{4/sqrt(2)});
  \draw (0,0,0) -- ({-4/sqrt(2)},{4/sqrt(2)},0);
  \draw (0,0,0) -- (0,{4/sqrt(2)},{4/sqrt(2)}) circle (0.07);

  \draw[fill=black] (4,0,0) circle (0.07);
  \draw[fill=black] (-4,0,0) circle (0.07);
  \draw[fill=black] (0,4,0) circle (0.07);
  \draw[fill=black] (0,-4,0) circle (0.07);
  \draw[fill=black] (0,0,-4) circle (0.07);
  \draw[fill=black] (0,0,4) circle (0.07);
  \draw[fill=black] ({4/sqrt(2)},0,{4/sqrt(2)}) circle (0.07);
  \draw[fill=black] (0,{4/sqrt(2)},{4/sqrt(2)}) circle (0.07);
  \draw[fill=black] ({-4/sqrt(2)},{4/sqrt(2)},0) circle (0.07);

  \begin{scope}[canvas is plane={O(0,0,0)x(0,0,1)y(-1/sqrt(2),1/sqrt(2),0)}]
    \draw[fill=red, opacity=0.2, line width=1] (4,0) arc (0:180:4);
        \draw[line width=1] (4,0) arc (0:180:4);
   \end{scope}

   \begin{scope}[canvas is plane={O(0,0,0)x(1,0,0)y(0,0,1)}]
      \draw[line width=1] (4,0) arc (0:360:4);
   \end{scope}

  %  \begin{scope}[canvas is plane={O(0,0,0)x(0,0,1)y(0,-1,0)}]
  %   \draw[orange, line width=1,fill=gray, opacity=0.2,] (4,0) arc (0:360:4);
  %   \draw[line width=1] (0,4) arc (90:-90:4);
  % \end{scope}

    \begin{scope}[canvas is plane={O(0,0,0)x(1,0,0)y(0,1,0)}]
      \draw[fill=gray, opacity=0.2, line width=1] (4,0) arc (0:360:4);
      \draw[line width=1] (4,0) arc (0:360:4);
   \end{scope}
  \begin{scope}[canvas is plane={O(0,0,0)x(0,0,1)y(-1/sqrt(2),1/sqrt(2),0)}]
    \draw[fill=red, opacity=0.2, line width=1] (4,0) arc (0:90:4) -- (0,0);
        \draw[line width=1] (4,0) arc (0:180:4);
   \end{scope}

   \begin{scope}[canvas is plane={O(0,0,0)x(1,0,0)y(0,0,1)}]
      \draw[fill=gray, opacity=0.3, line width=1] (4,0) arc (0:360:4);
      \draw[line width=1] (4,0) arc (0:180:4);
   \end{scope}

   \begin{scope}[canvas is plane={O(0,0,0)x({1/sqrt(2)},0, {1/sqrt(2)})y({-1/sqrt(6)},{1*sqrt(2/3)},{1/sqrt(6)})}]
     \draw[opacity=1, line width=1] (4,0) arc (0:120:4);
   \end{scope}

   \begin{scope}[canvas is plane={O(0,0,0)x(0,1,0)y(1/sqrt(2),0,1/sqrt(2))}]
     \draw[fill=yellow, opacity=0.3, line width=1] (4,0) arc (0:180:4);
     \draw[line width=1, name path=mahmut] (4,0,0) arc (0:180:4);

   \end{scope}

   \begin{scope}[canvas is plane={O(0,0,0)x(0,0,1)y(0,-1,0)}]
    \draw[black, line width=1,fill=gray, opacity=0.2,] (4,0) arc (0:360:4);
    \draw[line width=1] (0,4) arc (90:-90:4);

  \end{scope}

  \begin{scope}[canvas is plane={O(0,0,0)x(0,0,1)y(-1/sqrt(2),1/sqrt(2),0)}]
    \draw[line width=1] (4,0) arc (0:90:4);

   \end{scope}

   \begin{scope}[canvas is plane={O(0,0,0)x({1/sqrt(2)},0, {1/sqrt(2)})y({-1/sqrt(6)},{1*sqrt(2/3)},{1/sqrt(6)})}]
     \draw[fill=blue, opacity=0.3, line width =0] (4,0) arc (0:120:4) -- (0,0);

   \end{scope}
   
   \begin{scope}[canvas is plane={O(0,0,0)x(0,0,1)y(0,-1,0)}]
%    \draw[line width =0 ,fill=blue, opacity=1] (-45:4) arc (45:180:4) -- (0,0);
  \end{scope}

 \end{tikzpicture}

%%% Local Variables:
%%% mode: latex
%%% TeX-master: "../tropical"
%%% End: